
%
%
%
%
\input epsf
\input harvmac
%
%
%
%
%
%
%
\ifx\answ\bigans\else\output={
  \almostshipout{\leftline{\vbox{\pagebody\makefootline}}}\advancepageno
}
\fi
%
%
%
\def\mayer{\vbox{\sl\centerline{Department of Physics 0319}%
\centerline{University of California, San Diego}
\centerline{9500 Gilman Drive}
\centerline{La Jolla, CA 92093-0319}}}
%
%

%
%
\def\UCSD#1#2{\noindent#1\hfill #2%
\bigskip\supereject\global\hsize=\hsbody%
\footline={\hss\tenrm\folio\hss}}
%
%
\def\abstract#1{\centerline{\bf Abstract}\nobreak\medskip\nobreak\par #1}
%
%
%
%
\edef\tfontsize{ scaled\magstep3}
 \tfontsize  \tfontsize
 \tfontsize \font\titlei=cmmi10 \tfontsize
\font\titleis=cmmi7 \tfontsize \font\titleiss=cmmi5 \tfontsize
\font\titlesy=cmsy10 \tfontsize \font\titlesys=cmsy7 \tfontsize
\font\titlesyss=cmsy5 \tfontsize  \tfontsize
\skewchar\titlei='177 \skewchar\titleis='177 \skewchar\titleiss='177
\skewchar\titlesy='60 \skewchar\titlesys='60 \skewchar\titlesyss='60
%
%
%
%
%
\def\inv{^{\raise.15ex\hbox{${\scriptscriptstyle -}$}\kern-.05em 1}}
\def\lbar{{\lower.35ex\hbox{$\mathchar'26$}\mkern-10mu\lambda}} 

%
%
%
%
\def\slash#1{\rlap{$#1$}/} 
\def\dsl{\,\raise.15ex\hbox{/}\mkern-13.5mu D} 
\def\delsl{\raise.15ex\hbox{/}\kern-.57em\partial}
\def\Ksl{\hbox{/\kern-.6000em\rm K}}
\def\Asl{\hbox{/\kern-.6500em \rm A}}
\def\Dsl{\hbox{/\kern-.6000em\rm D}} 
\def\Qsl{\hbox{/\kern-.6000em\rm Q}}
\def\gradsl{\hbox{/\kern-.6500em$\nabla$}}
%
%
\def\lspace{\ifx\answ\bigans{}\else\qquad\fi}
\def\lbspace{\ifx\answ\bigans{}\else\hskip-.2in\fi} 
%
%
\def\boxeqn#1{\vcenter{\vbox{\hrule\hbox{\vrule\kern3pt\vbox{\kern3pt
        \hbox{${\displaystyle #1}$}\kern3pt}\kern3pt\vrule}\hrule}}}
%
%
\def\mbox#1#2{\vcenter{\hrule \hbox{\vrule height#2in
\kern#1in \vrule} \hrule}}
%
%
%
%
\def\CA{{\cal A}}  \def\CC{{\cal C}} 
   \def\CH{{\cal H}}
   
\def\CM{{\cal M}}  \def\CO{{\cal O}} \def\CP{{\cal P}}
\def\CQ{{\cal Q}} \def\CR{{\cal R}} \def\CS{{\cal S}} \def\CT{{\cal T}}

%
%
%
%
%

%

\def\bar#1{\overline{#1}}
\def\vev#1{\left\langle #1 \right\rangle}
\def\bra#1{\left\langle #1\right|}
\def\ket#1{\left| #1\right\rangle}
\def\abs#1{\left| #1\right|}

\def\darr#1{\raise1.5ex\hbox{$\leftrightarrow$}\mkern-16.5mu #1}

%
%
\def\frac#1#2{{\textstyle{#1\over #2}}} 
%
%
%
%

\def\Tr{\mathop{\rm Tr}}
\def\Im{\mathop{\rm Im}}

%
%
%
%

%
%
\def\ltap{\ \raise.3ex\hbox{$<$\kern-.75em\lower1ex\hbox{$\sim$}}\ }
\def\gtap{\ \raise.3ex\hbox{$>$\kern-.75em\lower1ex\hbox{$\sim$}}\ }
\def\gl{\ \raise.5ex\hbox{$>$}\kern-.8em\lower.5ex\hbox{$<$}\ }
\def\roughly#1{\raise.3ex\hbox{$#1$\kern-.75em\lower1ex\hbox{$\sim$}}}
%
%
\def\ie{\hbox{\it i.e.}}        \def\etc{\hbox{\it etc.}}
        
\def\etal{\hbox{\it et al.}}

\def\np#1#2#3{{Nucl. Phys. } B{#1} (#2) #3}
\def\pl#1#2#3{{Phys. Lett. } {#1}B (#2) #3}

\def\physrev#1#2#3{{Phys. Rev. } {#1} (#2) #3}
\def\ap#1#2#3{{Ann. Phys. } {#1} (#2) #3}

\relax

\def\lqcd{\Lambda_{\rm QCD}}
\def\wmunu{W_{\mu\nu}}
\def\itemsp#1{\medskip\item{\hbox to 1truecm{{#1\hfil}}}}
\def\cone{C^{(1)}_{j,n}}
\def\ctwo{C^{(2)}_{j,n}}
\def\cthree{C^{(3)}_{j,n}}

\def\cfour{C^{(4)}_{j,n}}

\def\boxtext#1{\vbox{\hrule\hbox{\vrule\kern5pt
       \vbox{\kern5pt{#1}\kern5pt}\kern5pt\vrule}\hrule}}
\def\tilde{\widetilde}
\noblackbox
%
%
%
%
%
%
\rightline{\vbox{\hbox{UCSD/PTH 92-10}\break\hbox{March 1992}}}
\bigskip
\centerline{{\titlefont{AN INTRODUCTION TO}}}
\bigskip
\centerline{{\titlefont{SPIN DEPENDENT DEEP INELASTIC
SCATTERING}}\footnote{*}{Lectures presented at the Lake Louise Winter
Institute,
February 1992}}
%
\bigskip\bigskip
\centerline{ANEESH V. MANOHAR}
\bigskip\mayer
\abstract{
The main
focus of
these lectures is on those aspects of deep inelastic
scattering that can
be derived directly from QCD using quantum field theory, without recourse to
phenomenological models.  The emphasis is on spin dependent scattering, but the
theory of spin averaged scattering is also discussed.  A
detailed analysis is given
for the case of spin 1/2 targets, with a brief discussion of higher spin
targets at the end. The QCD derivation of the Callan-Gross relation, the
longitudinal structure function $F_L$, and the Bjorken and Ellis-Jaffe sum
rules is presented. I also discuss the Wilczek-Wandzura contribution to $g_2$,
and why the Gottfried sum rule does not hold in QCD.
}

\UCSD{UCSD/PTH 92-10}{March 1992}
%

\newsec{Introduction}
These lectures provide an introductory account of the theory
of deep
inelastic scattering.  Many aspects
of deep
inelastic scattering can be analysed starting directly from
QCD, without
having to introduce phenomenological models. It is precisely
these
aspects of deep inelastic scattering that are most
important, and are the focus of these lectures. These are a set of
introductory
lectures for students, and are not intended to be a review. I have therefore
not given any references in the text. A few references are given at the end
which might be useful as an introduction to the vast literature on this
subject.

The outline of these lectures is as follows.
The kinematics of deep inelastic scattering is discussed
in Sec.~2.  Section~3 defines the leptonic tensor $\ell^{\mu\nu}$,
the hadronic tensor $W^{\mu\nu}$, and discusses the symmetry properties of
$W^{\mu\nu}$ and
its
decomposition into structure functions.
The idea of scaling is covered in Sec.~4.
The deep inelastic cross-section
is derived in Sec.~5 for spin-1/2 targets in terms of structure functions.
A useful inequality on  the structure function $g_1$
that follows from unitarity is derived in Sec.~6.
The
physical interpretation of the structure functions in terms of virtual photon
scattering is given
in Sec.~7. This section also discusses the Compton amplitude $T_{\mu\nu}$,
crossing symmetry, the optical theorem, and the analytic structure of the
scattering amplitude. A brief discussion of deep inelastic scattering in the
parton
model can be found in Sec.~8. The operator product
expansion and the derivation of the QCD sum rules for the
moments of the
structure functions are given  in Sec.~9. This section
is the most important part of these lectures, because
it covers the properties of deep inelastic scattering that can be {\sl derived}
directly from QCD.
Some applications of the results of Sec.~9, such as the Callan-Gross
relation, the longitudinal structure function, the
Ellis-Jaffe  and Bjorken sum
rules for
$g_1$, and the Wilczek-Wandzura contribution to $g_2$ are
discussed in Sec.~10. Anomalous dimensions, and
scaling violation are discussed in Sec.~11. Section~12
concludes with a brief summary of the structure functions
$b_1(x)$ and $\Delta(x)$ for spin one targets.

\newsec{The Kinematics of Deep Inelastic Scattering}

In a typical deep inelastic scattering
experiment, an incoming beam of leptons with energy $E$
scatters off a fixed hadronic
target. The energy and direction of the scattered lepton are
measured in
the detector, but the final hadronic state (usually denoted by
$X$) is
not measured experimentally. The  lepton interacts
with the
hadron target  through the exchange of a virtual photon; the target hadron
absorbs the
virtual photon, to produce the final state $X$. If the
target hadron
remains intact, the process is elastic scattering. The deep
inelastic
region is where the target hadron is blown apart by the
virtual photon,
and fragments into many particles. I will only discuss the
case of fixed-target deep inelastic scattering in detail. Deep inelastic
scattering
will soon be studied at HERA by colliding an electron beam
with a proton
beam. The kinematics for such colliding beam experiments is left as an
exercise for the
reader.

The basic diagram for deep inelastic scattering is show
schematically in \fig\exptfig{}.
There are numerous kinematic variables
which are used
in the discussion of deep inelastic scattering.
In the definitions given below, I will pick the $\hat z$
axis to be along the incident lepton
beam direction.
(Warning: in later
sections, I will pick the $\hat z$ axis to be along the
direction of the
virtual photon.) The kinematic variables are:
\midinsert
\vskip -1.5truecm
\epsffile{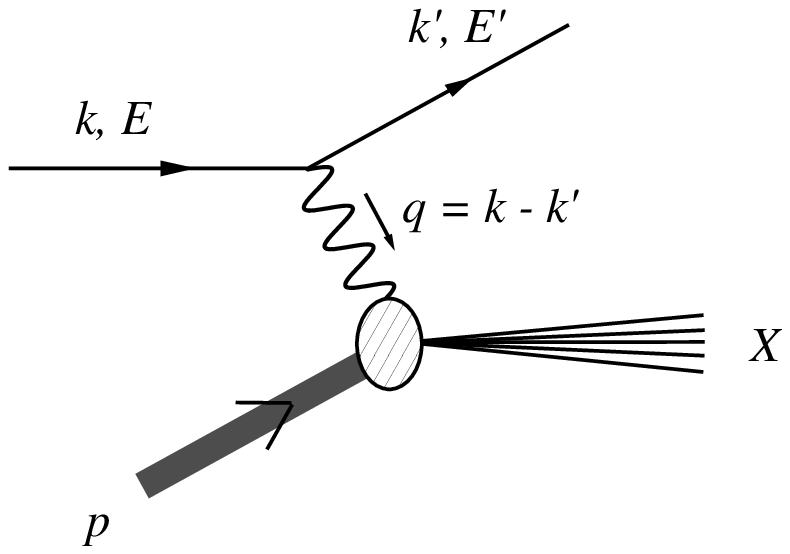}
\centerline{\tenrm FIGURE 1.}
{\tenrm \noindent The basic diagram for
deep inelastic lepton hadron scattering. The virtual photon
momentum is $\scriptstyle q$. The final hadronic state is not measured,
and is
denoted by $\scriptstyle X$.}
\endinsert

\bigskip\centerline{\underbar{{\sl Kinematic Variables}}}
\itemsp{$M$} The mass of the target hadron. The most
important case is for a proton or neutron target, in which
case $M$ is
the nucleon mass.
\itemsp{$E$} The energy of the incident lepton.
\itemsp{$k$} The  momentum of the initial lepton.
$k=(E,0,0,E)$, if the
lepton mass is neglected.
\itemsp{$\Omega$} The solid angle into which the outgoing
lepton is scattered.
\itemsp{$E'$} The energy of the scattered lepton.
\itemsp{$k'$} The momentum of the scattered
lepton,
\hfill\break
$k'=(E',E'\sin\theta\cos\phi,E'\sin\theta\sin\phi,E'
\cos\theta)$.
\itemsp{$p$} The momentum of the target, $p=(M,0,0,0)$, for a fixed
target experiment.
\itemsp{$q$} $=k-k'$, the momentum transfer in the
scattering
process, \ie\ the momentum of the virtual photon.
\itemsp{$\nu$} $=E-E'=p\cdot q/ M$, the energy loss of the
lepton.
\itemsp{$y$} $=\nu/E=p\cdot q/ p \cdot k$,
the fractional energy loss of the lepton.
\itemsp{$Q^2$} $=-q^2=2EE'(1-
\cos\theta)=4EE'\sin^2\theta/2$.
\itemsp{$x$} = $Q^2/2 M \nu=Q^2/2 p \cdot q = Q^2/2 M E y$.
\itemsp{$\omega$} = $1/x$.
\medskip
The variable $x$ was first introduced by Bjorken, and is
crucial to
understanding deep inelastic scattering because we will
see that
QCD predicts that  structure functions are functions of
$x$ and
independent of $Q^2$ to leading order, a property
known as scaling.
Higher order corrections in QCD produce a small logarithmic
$Q^2$
dependence of the structure functions which is calculable
for large
$Q^2$ since QCD is asymptotically free. I will therefore use the
following definition of deep inelastic scattering:
\bigskip
\centerline{{\boxtext{\sl
\hbox to 5 truein{\hfill Deep
inelastic scattering is the study of lepton-hadron
scattering\hfill }
\hbox to 5 truein {\hfill in the
limit that $x$ is fixed, and $Q^2\rightarrow\infty$.\hfill}
}}}
\bigskip
\noindent
The invariant mass of the final hadronic system $X$ is
\eqn\iione{
M^2_X = (p+q)^2
= M^2 + 2 p \cdot q + q^2.
}
The invariant mass of $X$ must be at least
that of a nucleon, since baryon number is conserved in the
scattering process. This gives the inequality
\eqn\invmass{
M^2_X \ge M^2\quad \Rightarrow\quad M^2 + 2 p \cdot q - Q^2 \ge M^2
\quad\Rightarrow\quad x \le 1.
}
Since $Q^2$ and $\nu$ are both positive, $x$ must also be
positive.
The lepton energy loss $E-E'$ must be between zero and $E$,
so
the physically allowed kinematic region is
\eqn\iitwo{
0 \le x \le 1,\qquad 0 \le y \le 1.
}
Eq.~\invmass\ can be written in the form
\eqn\masstwo{
x={Q^2\over 2 p\cdot q} = 1- {M_X^2-M^2\over 2 p\cdot q}.
}
The value $x=1$ implies that $M^2_X=M^2$, and so $x=1$
corresponds to
elastic scattering.
 Any
fixed
hadron state $X$ with invariant mass $M^2_X$ contributes to
the
cross-section at the value of $x$ obtained from
Eq.~\masstwo,
\eqn\xvalue{
x = {1\over 1 + \left(M^2_X-M^2\right)/Q^2}.
}
In the deep inelastic limit $Q^2\rightarrow\infty$, so that
any state $X$ with
fixed mass
$M_X$ gets driven to $x=1$. In particular, all nucleon
resonances such
as the $N^*$ get pushed to $x=1$. The invariant mass of the
hadronic
state that contributes to the cross-section at a given value
of $Q^2$
increases as $x\rightarrow0$. For a fixed
value of $Q^2$,
there is a small region around $x=1$ of width $\lqcd^2/Q^2$
which probes hadronic resonances with masses around that of
the
nucleon. This resonance region is present in any real experiment since $Q^2$ is
finite, but is not present in the formal deep inelastic limit.
Outside the resonance region, the invariant mass $M_X^2$ is of order $Q^2$.

The experimental measurements give the cross-section as a
function of
the final lepton energy and scattering angle,
${d^2\sigma/dE' d\Omega}$.
The results are often presented instead by giving the
differential cross-section as a function of $(x,\,Q^2,\,\phi)$ or
$(x,\,y,\,\phi)$. The Jacobian for
converting between these cases is easily worked out using
the
definitions of the kinematic variables,
\eqn\iijac{\eqalign{
{\partial(x,Q^2)\over \partial(x,y)} &= \left|
\matrix{1&0\cr 2 M E y&2 M Ex\cr}\right| = 2 M E x =
{Q^2\over y},\cr
\noalign{\medskip}
{\partial(x,y)\over \partial(E',\cos\theta)} &= \left|
\matrix{?&{-2EE'\over2M\nu}\cr-{1\over E}&0\cr}\right| =
{E'\over M\nu}.
}}
Thus the cross-sections are related by
\eqn\convert{
{d^2\sigma\over dx\, dy\, d\phi} = \left({M \nu\over
E'}\right)
{d^2\sigma\over dE' d\Omega} = \left({Q^2\over y}\right)
{d^2\sigma\over dx\, dQ^2 d\phi}.
}
For most cases of interest, the cross-section will be
independent
of the azimuthal angle $\phi$. In these lectures, I will
give the
differential cross-section in ($x$, $y$, $\phi$). The
variables $x$ and
$y$ are Lorentz invariant, because they are defined using
the inner
products of four vectors. Thus the formula for the
differential
cross-section as a function of $x$, $y$, and $\phi$ is also valid for
the
kinematics at HERA, since cross-sections  (and $\phi$) are invariant under
boosts
along the collision axis. The only difference is that the
transformation formula
between ($x$, $y$), and ($E'$, $\theta$) is modified.

\midinsert
\epsffile{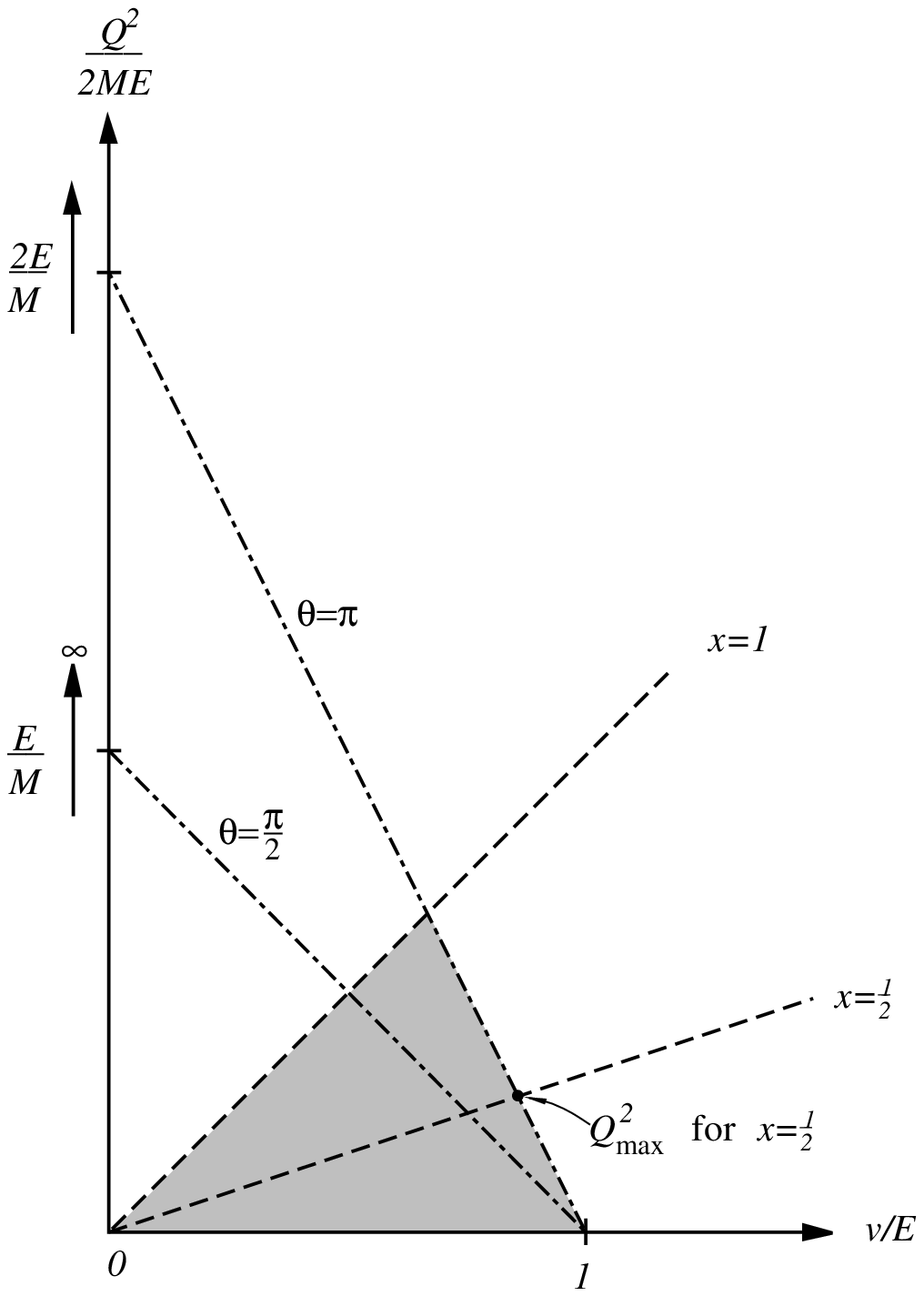}
\centerline{\tenrm FIGURE 2.}
{\tenrm \noindent The kinematically allowed region for deep
inelastic
scattering is shown in grey. The dot-dashed lines are lines of constant
scattering angle $\scriptstyle \theta$, and
the dashed lines are lines of constant $\scriptstyle x$. In the deep inelastic
limit, the intercept of the constant $\scriptstyle \theta$ lines with the
vertical axis becomes infinite.}
\endinsert
There are three independent variables which describe the
kinematics, $E'$, $\theta$ and $\phi$.
The dependence on $\phi$ is trivial, so I will only  discuss
the
dependence on the other two independent variables. It is
convenient to plot the allowed kinematic region in the
$(Q^2/2ME)-(\nu/E)(=y)$
plane, as shown
in \fig\dalitz{}. The boundary of
the
physical region is given by the requirements that
\eqn\boundary{
0 \le \theta \le \pi,\quad 0\le\nu\le E,\quad 0\le x \le 1.
}
Since $x=Q^2/2M\nu=(Q^2/2ME)/(\nu/E)$, the contours of
constant $x$ are
straight lines through the origin with slope $x$. The relation
between
$Q^2$ and $\theta$ can be written as
\eqn\iifour{
Q^2 = 2 E E'(1-\cos\theta)\quad \Rightarrow\quad {Q^2\over 2 M E } =
{ 1 \over M}
(E-\nu)(1-\cos\theta).
}
Thus lines of constant $\theta$ are straight lines passing
through the point $\nu/E=1$, and intersecting the $Q^2/2ME$ axis
at $Q^2/2ME =
(E/M)(1-\cos\theta)$. Lines of fixed $\theta$ become
steeper
as the beam energy increases, whereas lines of fixed $x$ remain constant.

The $Q^2$
dependence of
the kinematic variables is crucial to understanding which
terms are
important in the deep inelastic limit. A generic point in
the kinematic
plane is given by some value of $x$ and $y$. As
$E\rightarrow\infty$ for
a fixed value of $x$ and $y$, the variables $\nu/E$ and
$Q^2/2ME$ are
fixed. This implies that in the deep inelastic limit,
$\nu$ and $E$ are each proportional
to $Q^2$,
\eqn\iifive{
\nu\ \propto\ E\ \propto\ {Q^2\over M}.
}
A generic point in the physical region has $(1-\cos\theta)$ of
order $M/E$,
\eqn\iisix{
\left(1 - \cos\theta\right)\ \propto\ {M\over E}\ \propto\ {M^2\over Q^2},
}
so that the scattering angle $\theta$ is of order
$M/Q$,
\eqn\iiseven{
\theta\ \propto\ {M\over Q},
}
and goes to zero as $Q^2\rightarrow\infty$.

The allowed region in \dalitz\ also shows that for fixed
beam energy
$E$,  there is a limit to the $Q^2-x$ region which can be
explored experimentally.
The
small $x$ region is also the small $Q^2$ region, because
lines of
constant $x$ approach the horizontal axis for small $x$.
For a fixed
value of
$x$, the maximum allowed value of $Q^2$ is at the
intersection
of the
line $\theta=\pi$ with the line for fixed $x$. It
is elementary to find the intersection point of the two
lines,
\eqn\qmax{
Q^2_{\rm max} = 2 M E x\left( {2 E \over 2 E + M x}
\right) \approx 2 M E x, \qquad (E\gg M x).
}
To be in the deep inelastic region, one needs $Q^2$ to be
larger than a
few $(\rm GeV)^2$, so this places a limit on the smallest
value of $x$
accessible for a given beam energy. For example, with a
500~GeV lepton
beam, and assuming $Q^2\ge 10\ (\rm GeV)^2$ is large enough
to be considered deep inelastic scattering, the smallest
measurable value of $x$ is $10^{-2}$. The factor of $2ME$ in Eq.~\qmax\
is
proportional to the centre of mass energy squared, \ie\ to
$s=(p+k)^2$.
HERA will collide 28~GeV leptons on 820~GeV protons, so the
value of $s$
is $4 (28) (820)\ {(\rm GeV)^2}$. This will allow the
measurement of
structure functions down to $x\approx10^{-4}$. A similar
measurement in
a fixed target experiment would require a 50~TeV lepton
beam.

Finally, it is useful to have formul\ae\ for the  different
components
of $q$ as a function of $x$ and $y$. The result is obtained
by
substituting for $\theta$ in terms of $Q^2$ into the
definition of $q$,
\eqn\iieight{
q = k-k'=(E-E',E'\sin\theta\cos\phi,E'\sin\theta\sin\phi,E-
E'\cos\theta),
}
\eqn\qcomps{\eqalign{
&\qquad q^0={Q^2\over 2 M x y},\qquad q^3={Q^2\over 2 M x y}+ M x
y,\qquad\cr
&\quad q^\perp=Q\sqrt{1-y}\sqrt{1-{M^2 x^2 y^2\over Q^2 (1-
y)}}\approx Q\sqrt{1-y}.\cr
}}
Note that $q^\perp$ is of order $Q$, whereas $q^0$ and $q^3$
are both of
order $Q^2/M$. Also, $q^0$ and $q^3$ are nearly equal, with
$q^0-q^3$
being of order $M$. These properties of $Q$ will be useful
when we
compute cross-sections in the deep inelastic limit.

\newsec{Electroproduction, $\ell_{\mu\nu}$, and
$W_{\mu\nu}$}

The basic Feynman graph for deep inelastic scattering is
\exptfig. The scattering amplitude $\CM$ is given by
\eqn\scatamp{
i\CM = (-ie)^2 \left({-i g_{\mu\nu}\over q^2}\right)
\bra{k'}
j^\mu_\ell(0) \ket{k,s_\ell}
\bra{X} j^\nu_h(0) \ket{p,\lambda},
}
where $s_\ell$ is the polarisation of the initial lepton,
$\lambda$ is the polarisation of the initial hadron, and
$j^\mu_\ell$, $j^\mu_h$ are the leptonic and hadronic
electromagnetic currents respectively. The coupling constant
$e$ is not
included in the definition of the current, and hence has
been included
explicitly in $\CM$. The quantity $\lambda$ which describes
the hadron
polarisation can be chosen to be the value of the spin along
an
arbitrary quantisation axis (usually chosen to be the beam
direction).
For a spin 1/2 target, $\lambda=\pm1/2$. We will discuss the
hadron
polarisation in greater detail later in this section.

The differential scattering cross-section is obtained from
$\CM$ by
squaring, and multiplying by the phase space factors.
The polarisation of the final lepton and hadron states
are not measured, and have to be
summed over. This gives the differential cross-section
\eqn\cross{\eqalign{
d\sigma &= \sum_X \int {d^3 k'\over (2\pi)^3 2 E'}
\ (2\pi)^4 \delta^4(k+p-k'-p_X)\ {\abs{\CM}^2 \over (2 E)(2 M)
(v_{\rm rel}=1)},\cr
\noalign{\smallskip}
&=\sum_X
\int {d^3 k'\over (2\pi)^3 2 E'}
\ {(2\pi)^4 \delta^4(k+p-k'-p_X) \over (2 E)(2 M)}
\ {e^4\over Q^4} \cr & \times \bra{p,\lambda}
 j^\mu_h(0) \ket{X}\bra{X} j^\nu_h(0) \ket{p,\lambda}
\bra{k,s_\ell} j_{\ell\mu}(0) \ket{k'} \bra{k'}
j_{\ell\nu}(0)
\ket{k,s_\ell},\cr
}}
where I have used the relation
\eqn\iiione{
\bra{\alpha} j^\mu \ket{\beta}^* = \bra{\beta} j^\mu
\ket{\alpha},
}
which is true because the current is hermitian,
$j_\mu^\dagger=j_\mu$.
It is conventional to define the leptonic tensor
$\ell_{\mu\nu}$ by
\eqn\lmunu{
\ell^{\mu\nu} = \sum_{\rm final\ spin}\bra{k'} j^\nu_\ell(0)
\ket{k,s_\ell}
\bra{k,s_\ell} j^\mu_\ell(0) \ket{k'},
}
so that the leptonic current matrix elements in Eq.~\cross\
can be
replaced by $\ell_{\mu\nu}$.
The definition of the hadronic tensor is slightly more
complicated. The tensor $W_{\mu\nu}$ is defined by
\eqn\wmunudef{
W^{\mu\nu} (p,q)_{\lambda'\lambda} = {1\over 4\pi} \int
d^4x\,e^{iq\cdot x}
\bra{ p,\lambda'} [j^\mu(x), j^\nu (0)]\ket{p,\lambda}\ ,
}
where $\lambda$ and $\lambda'$ are the polarisations of the
initial and final hadron.
Inserting a complete set of states
gives
\eqn\expandwmunu{\eqalign{
W^{\mu\nu} (p,q)_{\lambda'\lambda} = {1\over 4\pi} \sum_X
\int d^4x\,e^{iq\cdot x}
&\Bigl[\bra{ p,\lambda'} j^\mu(x)\ket{X}\bra{X} j^\nu
(0)\ket{p,\lambda}\cr
-&\bra{ p,\lambda'} j^\nu(0)\ket{X}\bra{X} j^\mu
(x)\ket{p,\lambda}\Bigr],\cr
}}
where the sum on $X$ is a sum over the allowed phase space
for the final
state $X$.
Translation invariance implies that
\eqn\trans{\eqalign{
\bra{ p,\lambda'} j^\mu(x)\ket{X} &= \bra{ p,\lambda'}
j^\mu(0)\ket{X}
e^{i(p-p_X)\cdot x},\cr
\noalign{\smallskip}
\bra{X} j^\mu (x)\ket{p,\lambda}&=\bra{X} j^\mu
(0)\ket{p,\lambda}
e^{i(p_X-p)\cdot x}.\cr
}}
Inserting Eq.~\trans\ into Eq.~\expandwmunu\ gives
\eqn\wmunutwo{\eqalign{
W^{\mu\nu} (p,q)_{\lambda'\lambda} = {1\over 4\pi} \sum_X
&\Bigl[(2\pi)^4 \delta^4(q+p-p_X) \bra{ p,\lambda'}
j^\mu(0)\ket{X}\bra{X} j^\nu (0)\ket{p,\lambda}\cr -
&(2\pi)^4 \delta^4(q+p_X-p)
\bra{ p,\lambda'} j^\nu(0)\ket{X}\bra{X} j^\mu
(0)\ket{p,\lambda}\Bigr].\cr
}}
The only allowed final states are those with $p_X^0\ge p^0$,
since
$M^2_X \ge M^2$. Since $q^0 > 0$, only the first delta
function in
Eq.~\wmunutwo\ can be satisfied, and the sum in $W_{\mu\nu}$
reduces to the expression in Eq.~\cross\ involving the
hadronic
currents, and the energy-momentum conserving delta
function (up to a
factor of $1/4\pi$). Only the first term in
Eq.~\expandwmunu\ contributes,
which is equivalent to the statement that one could have
defined
$W_{\mu\nu}$ in Eq.~\wmunudef\ simply as the matrix element of
$j^\mu(x) j^\nu(0)$
without the commutator. The reason for using the commutator
is that then
$W_{\mu\nu}$ has a nicer analytic structure when continued
away from the
physical region which we will use extensively in Sec.~9.
Substituting
for $\ell_{\mu\nu}$ and $W^{\mu\nu}$ in Eq.~\cross\ gives
\eqn\crosstwo{
d\sigma = {e^4\over Q^4}\int {d^3 k'\over (2\pi)^3 2 E'}
\ {4 \pi\ \ell^{\mu\nu}\ W^{\mu\nu}(p,k-k')_{\lambda\lambda}
\over (2 E)(2 M) (v_{\rm rel}=1)},
}
so that
\eqn\finalcross{\eqalign{
{d^2\sigma\over dE' d\Omega} &= {e^4\over 16 \pi^2
Q^4}\left({E'\over M E}\right) \ell_{\mu\nu}
\ W^{\mu\nu}(p,q)_{\lambda\lambda},\cr
\noalign{\smallskip}
{d^2\sigma\over dx\, dy\, d\phi} &= {e^4\over 16 \pi^2 Q^4}\
y
\ \ell_{\mu\nu}\ W^{\mu\nu}(p,q)_{\lambda\lambda}.
}}
All the information about the deep inelastic
scattering cross-section is contained in the leptonic and
hadronic tensors
$\ell_{\mu\nu}$ and $\wmunu$.

\bigskip
\leftline{{\it \underbar{The Leptonic Tensor $\ell^{\mu\nu}$}}}
\nobreak\medskip\nobreak

The leptonic tensor is trivial to compute since the leptons
are pointlike fermions,
\eqn\lmunuone{
\ell^{\mu\nu} = \sum_{\rm final\ spin} \bar u (k')
\,\gamma^\nu\, u(k,s_\ell)\
\bar u (k,s_\ell)\, \gamma^\mu\, u(k').
}
The sum over final state spinors uses the identity
\eqn\finalsum{
\sum_{\rm final\ spin} u(k')\, \bar u(k') = \slash k' + m,
}
where the spinors are normalised to $2E$.

The polarisation of a spin 1/2 particle can be described by
a spin
vector $s_\ell$, defined in the rest frame of the particle by
\eqn\spindefone{
2 \vec s_\ell = \bar u(k,s_\ell)\,\vec \sigma\, u(k,s_\ell) = \bar u(k,s_\ell)
\, \vec \gamma\, \gamma_5\, u(k,s_\ell),
}
where the states have the conventional relativistic norm of
$2m$ in the
rest frame. The vector $s^\mu_\ell$ is defined in an arbitrary
Lorentz frame
by boosting $s_\ell=(0,\vec s_\ell)$ from the rest frame, or
equivalently, by
\eqn\spindef{
2 s^\mu_\ell = \bar u(k,s_\ell)\, \gamma^\mu \gamma_5\, u(k,s_\ell).
}
For a spin-1/2 particle at rest with spin up along the $\hat
z$ axis,
the spin vector is $\vec s_\ell = m \hat z$. This differs from
the
conventional normalisation of $s_\ell$ by a factor of the fermion
mass $m$.
It is extremely useful to use this normalisation for the
spin to avoid
unnecessary factors of $m$ appearing in the spin dependent
cross-section. In the extreme relativistic limit, all mass
effects
can be neglected in the calculation, and all formul\ae\
should be
written in terms of relativistic spinors normalised to $2E$,
without any
additional factors of $m$. This suggests that one define
spin by
Eq.~\spindef\ without any factors of $m$, and thus define a
spin vector
with the dimensions of mass, which is $m$ times the
conventional
definition. With this normalisation for $s_\ell$,
longitudinally polarised fermions in the extreme
relativistic limit have $s_\ell=\CH_\ell k$, where $k$ is
the lepton
momentum and $\CH_\ell=\pm$ is the lepton helicity.

The initial state spinor product can be written in terms of
the standard
spin projection operator,
\eqn\spinsum{
u(k,s_\ell)\,\bar u (k,s_\ell)= (\slash k + m)\ {1 + \gamma_5
\slash s_\ell/m_\ell \over 2 }.
}
Substituting Eq.~\finalsum\ and \spinsum\ into Eq.~\lmunuone\
gives
\eqn\lmunu{\eqalign{
\ell^{\mu\nu} &= \Tr\ (\slash k' +
m_\ell)\, \gamma^\nu\, (\slash k + m_\ell)\ {1 + \gamma_5 \slash
s_\ell/m_\ell \over 2 }\ \gamma^\mu,\cr
&=2 \left( k^\mu k'^\nu + k^\nu k'^\mu - g^{\mu\nu} (k\cdot
k' - m^2_\ell)
- i \epsilon^{\mu\nu\alpha\beta} q_\alpha
s_{\ell\beta}\right),\cr
&\approx 2 \left( k^\mu k'^\nu + k^\nu k'^\mu - g^{\mu\nu}
k\cdot
k'
- i \epsilon^{\mu\nu\alpha\beta} q_\alpha
s_{\ell\beta}\right),\cr
}}
neglecting the lepton mass. There are no factors of $m_\ell$
in the spin
term with our normalisation convention Eq.~\spindef.
Note that the spin-independent part
of the leptonic tensor is symmetric in $\mu\nu$, and the
spin-dependent part is antisymmetric in $\mu\nu$. Thus an
unpolarised lepton beam probes only the symmetric part of
$W_{\mu\nu}$, and the polarisation asymmetry in the cross-sections
probes only the antisymmetric
part
of $W_{\mu\nu}$.

\bigskip
\leftline{{\it \underbar{The Hadronic Tensor $W^{\mu\nu}$ for Spin-1/2
Targets}}}
\nobreak\medskip\nobreak

The structure of the hadron relevant for deep inelastic
scattering can be completely characterised by the hadronic
tensor $W_{\mu\nu}$. Unlike the leptonic tensor which is
known, $W_{\mu\nu}$ cannot be computed directly from QCD
because of non-perturbative effects in the strong
interactions. The strong interactions are invariant under
parity and time reversal, and these symmetries place
restrictions on the form of $W_{\mu\nu}$. In this section,
I will concentrate on hadronic targets with spin-1/2, which
is the most important case experimentally.
The tensor $W_{\mu\nu}$ depends on the polarisations
$\lambda,\lambda'$, which for a spin-1/2 target can each
take the values $\pm1/2$. We will need to discuss the case
where
$\lambda\not=\lambda'$ to describe an arbitrary target
polarisation,
even though it appears that only
$\lambda=\lambda'$ occurs in Eq.~\finalcross.

The most general polarisation
state of a spin $J$ target in its rest frame can be written
as a
density matrix
\eqn\pol{
\rho = \sum_{\lambda=-J}^J
\ \sum_{\lambda'=-J}^J
\ket{\lambda}\rho_{\lambda\lambda'}\bra{\lambda'}.
}
The states $\ket{\lambda}$ and $\ket{\lambda'}$ transform
as the spin-$J$ representation under the rotation group.
Thus the density matrix $\rho$ transforms under the
representation $J\otimes J =0\oplus1\oplus\ldots \oplus2J$. The density
matrix of a
spin-$J$ target can thus be described by $2J+1$ irreducible
tensors.
We
will restrict ourselves in this section to
$J=1/2$, whose
polarisation density matrix can
be decomposed into irreducible tensors of spin 0 and 1, \ie\
by a scalar
and a vector. It is easy to see that one can define a
scalar and a vector from $\rho$ by
\eqn\polirred{
\rho_{\rm scalar} \equiv \Tr \rho =1,\qquad
\vec\rho_{\rm vector} \equiv \Tr \rho\,\vec\sigma = \vec
s_h / M ,
}
which defines $\vec s_h$.
$\rho$ can be expressed in terms of these irreducible tensors,
\eqn\poldecomp{
\rho = {1\over2}\left( 1 + \vec \sigma\cdot {\vec s_h\over
M}\right).
}
The only complication for targets with spin greater than 1/2
is that the
decompositions analogous to Eqs.~\polirred\ and \poldecomp\ are
more
complicated, which is why I have restricted the analysis of
$W_{\mu\nu}$
to spin-1/2 targets.
A spin-1/2 polarised target can be described by an
axial vector $s_h$, which  has been defined with
an additional factor of the target mass $M$ for the same
reasons that we absorbed a factor of the leptonic mass into
the lepton polarisation vector $s_\ell$. The magnitude of
$\vec s_h$ depends on the degree of polarisation of the
target;
for an unpolarised target, $\vec s_h=0$, for a 100\%
polarised target,
$\abs{\vec s_h}=M$, and for a target with fractional
polarisation $f$,
$\abs{\vec s_h}=f M$. The four vector $s_h^\mu$ is defined to be
$(0,\vec s_h)$ in the rest frame of the target, and is defined
in other reference frames by a Lorentz boost.

The tensor $W_{\mu\nu}(p,q,s_h)$ for a spin-1/2 target is
defined by
\eqn\wspinhalf{
W_{\mu\nu}(p,q,s_h) = \sum_{\lambda,\lambda'}
\rho_{\lambda\lambda'}\,
W_{\mu\nu}(p,q)_{\lambda'\lambda} = \Tr \rho\, W_{\mu\nu}
}
treating $W_{\mu\nu}$ as a matrix in spin space with labels
$\lambda,\lambda'$. Thus the most general tensor
$\wmunu$ for a
spin-1/2 target can be written as a function of the
vectors $p$,
$q$ and $s_h$. Note that $s_h$ contains the spin information
of both the
initial and final state spinors through the decomposition of
the density
matrix Eq.~\poldecomp. Thus $s_h$ can only occur {\sl
linearly} in
$\wmunu$; there are no terms with two factors of $s_h$.
\bigskip
\leftline{{\it \underbar{The Decomposition of $W^{\mu\nu}$ into Structure
Functions}}}
\nobreak\medskip\nobreak
\indent The deep inelastic structure functions are obtained by
writing down the
most general tensor decomposition of $\wmunu$.
The strong interactions are invariant under parity.
The parity transform of the electromagnetic current is
\eqn\xxyyone{
\CP j^\mu(x) \CP^{-1}= j^{\mu_P}(x_P),
}
with $\mu_P = \mu$
for $\mu=0$, and
$\mu_P=-\mu$ for $\mu=1,2,3$. To avoid writing complicated
formul\ae, I
will use the obvious convention $j^{-1}=-j^1$, \etc, for the
transformation of the various components of a vector under
parity.
The parity transform of the point $x$ is
$x_P=(t,-\vec x)$, and the parity transform of the
polarisation density matrix is
\eqn\rhoparity{
\CP \rho \CP^{-1} = \rho_P,\qquad \Tr \rho_P =1,\qquad
M \Tr \rho_P\vec\sigma = \vec s_h,
}
since $s_h$ is an axial vector.
This
implies
\eqn\parityinvone{\eqalign{
W^{\mu\nu} (p,q,s_h) &= {1\over 4\pi} \int d^4x\,e^{iq\cdot
x}
\Tr \rho\, [j^\mu(x), j^\nu (0)]\cr
\noalign{\smallskip}
&= {1\over 4\pi} \int d^4x\,e^{iq\cdot x}
\Tr\, \CP \rho \CP^{-1}\ \CP [j^\mu(x), j^\nu (0)]\CP^{-1}\cr
\noalign{\smallskip}&= {1\over 4\pi} \int d^4x\,e^{iq\cdot x}
\Tr \rho_P\, [j^{\mu_P}(x_P), j^{\nu_P} (0)].\cr
}}
Defining $q_P=(q^0,-\vec
q)$,
$(s_h)_P = (-s_h^0,\vec s_h)$, and using
$q_P\cdot x_P = q \cdot x$, one can rewrite the last
equality in Eq.~\parityinvone\ as
\eqn\parityinv{
W^{\mu\nu} (p,q,s_h) = W^{\mu_P\nu_P} (p_P,q_P,(s_h)_P),
}
which is the condition that $\wmunu$ is parity invariant.

Invariance under time-reversal is only slightly more
complicated, because time reversal is antiunitary.
The time-reversal of momentum is defined by $p_T=(p^0,-\vec
p)$, spin by
$(s_h)_T = (s^0_h,-\vec s_h)$, and coordinates by
$x_T=(-t,\vec x)$. Note that time
reversal takes $x^0$ to $-x^0$, but still takes $p^0$ to
$+p^0$, \ie\
time reversal changes the sign of time, but not of energy.
Time reversal invariance then implies that
\eqn\tinvone{\eqalign{
W^{\mu\nu} (p,q,s_h) &= {1\over 4\pi} \int d^4x\,e^{iq\cdot
x}
\Tr \rho\, [j^\mu(x), j^\nu (0)]\cr
\noalign{\smallskip}&= {1\over 4\pi} \int d^4x\,e^{iq\cdot x}
\left( \Tr \CT \rho \CT^{-1}\ \CT [j^\mu(x), j^\nu
(0)]\CT^{-1}\right)^*\cr
\noalign{\smallskip}&= {1\over 4\pi} \int d^4x\,e^{iq\cdot x}
\left( \Tr \rho_T\, [j^{\mu_T}(x_T), j^{\nu_T}
(0)]\right)^*,\cr
}}
where the complex conjugation in the second line is present
because time
reversal is antiunitary. Using the identity $q \cdot x = -
q_T \cdot
x_T$, we get
\eqn\vvone{
e^{i q \cdot x} = \left( e^{i q_T\cdot x_T}\right)^*,
}
so that time reversal invariance implies that
\eqn\tinv{
W^{\mu\nu}(p,q,s_h) = W^{\mu_T\nu_T}(p_T,q_T,(s_h)_T)^*.
}

The electromagnetic currents and density matrix are
hermitian, which implies
\eqn\herminv{\eqalign{
W^{\mu\nu}(p,q,s_h)^* &= {1\over 4\pi} \int d^4x\,e^{-
iq\cdot x}
\Tr \rho^\dagger\, [j^\mu(x), j^\nu (0)]^\dagger\cr
\noalign{\smallskip}&= {1\over 4\pi} \int d^4x\,e^{-iq\cdot x}
\Tr \rho\, [j^\nu(0), j^\mu (x)]\cr
\noalign{\smallskip}&= {1\over 4\pi} \int d^4x\,e^{-iq\cdot x}
\Tr \rho \,[j^\nu(-x), j^\mu (0)] = W^{\nu\mu}(p,q,s_h),
}}
where the first line follows from the  identity
$(\Tr A B)^*= \Tr (A B)^\dagger$ and the last line follows from
translation
invariance. Note that all three identities,
Eqs.~\parityinv, \tinv, and \herminv\ relate $W^{\mu\nu}$ for
$q^0$ to a transformed $W^{\mu\nu}$ at the same value of
$q^0$.

There is a final identity that relates $\wmunu$ for $q^0>0$
to $\wmunu$
for $q^0<0$ which follows from the antisymmetry of the
commutator and translation invariance:
\eqn\crossinv{\eqalign{
W^{\mu\nu}(p,q,s_h) &= -{1\over 4\pi} \int d^4x\,e^{iq\cdot
x}
\Tr \rho\, [j^\nu(0), j^\mu (x)]\cr
&= -{1\over 4\pi} \int d^4x\,e^{iq\cdot x}
\Tr \rho\, [j^\nu(-x), j^\mu (0)]\cr
&= -{1\over 4\pi} \int d^4x\,e^{-iq\cdot (-x)}
\Tr \rho\, [j^\nu(-x), j^\mu (0)] = -W^{\nu\mu}(p,-q,s_h).
}}
This identity is the photon crossing symmetry identity for $\wmunu$, since
exchanging $\mu\nu$ and letting $q\rightarrow-q$ is equivalent to exchanging
the outgoing and incoming photons.
Finally current conservation, $\partial_\mu j^\mu(x)=0$, implies that
\eqn\currcons{
q_\mu W^{\mu\nu}(p,q,s_h) = q_\nu W^{\mu\nu}(p,q,s_h) = 0,
}
which greatly restricts the possible tensor structures
allowed for
$\wmunu$.

The most general hadronic tensor for polarised deep
inelastic scattering from  spin-1/2 targets can be written
in terms of $p,q,s_h$ and the invariant tensors
$g_{\alpha\beta}$ and $\epsilon_{\alpha\beta\lambda\sigma}$
with the constraints
Eqs.~\parityinv--\currcons. The algebra is
straightforward but tedious. It is conventional to write
$\wmunu$ in the form
\eqn\wmunuhalflong{
\eqalign{W_{\mu\nu} &= F_1 \left(-g_{\mu\nu} + {q_\mu
q_\nu\over q^2}\right)
+ {F_2\over p \cdot q} \left(p_\mu - {p\cdot q \ q_\mu\over
q^2}\right)
\left(p_\nu - {p\cdot q\ q_\nu\over q^2}\right)\cr
\noalign{\smallskip}&+ {ig_1\over p\cdot q}\ \epsilon_{\mu\nu\lambda\sigma}
q^\lambda s^\sigma +
{ig_2\over (p\cdot q)^2}\ \epsilon_{\mu\nu\lambda\sigma}
q^\lambda \left( p\cdot q
\, s^\sigma - s\cdot q\,p^\sigma\right)\ , \cr}
}
where the coefficients of the different tensor structures
are called structure functions. The leptonic current is also
conserved, so that $q^\mu \ell_{\mu\nu}= q^\nu
\ell_{\mu\nu}=0$. Thus it is convenient to simplify the
expression for $\wmunu$ by omitting all $q_\mu$ and
$q_\nu$ terms before contracting with $\ell_{\mu\nu}$,
\eqn\wmunuhalf{
\eqalign{W_{\mu\nu} &= - F_1\ g_{\mu\nu}
+ {F_2\over p\cdot q}\ p_\mu
p_\nu + {ig_1\over \nu}\ \epsilon_{\mu\nu\lambda\sigma} q^\lambda
s^\sigma_h \cr &\qquad +
{ig_2\over \nu^2}\ \epsilon_{\mu\nu\lambda\sigma} q^\lambda
\left( p\cdot q
\, s_h^\sigma - s_h\cdot q\,p^\sigma\right)\ , \cr}
}
(Alternatively, one can drop the $q_\mu$ and $q_\nu$ terms
in $\ell_{\mu\nu}$. However, it is incorrect to drop the
$q_\mu$ and $q_\nu$ terms in both $\ell_{\mu\nu}$ and
$\wmunu$!)
There are also structure functions $W_1$, $W_2$,
$G_1$
and $G_2$ proportional to $F_1$, $F_2$, $g_1$ and
$g_2$ which
are sometimes used in the literature,
\eqn\xxyytwo{\vcenter{\openup1\jot
\halign{$\hfil#$&${}#\hfil$
      &\qquad$\hfil#$&${}#\hfil$\cr
M W_1 &= F_1,&\nu W_2 &= F_2,\cr
M^2\nu G_1 &= g_1,& M \nu^2 G_2 &= g_2,\cr}}}
but I will never refer to these, because they do not scale
in the deep
inelastic limit.

For a spin-1/2 target, the symmetric
part of $\wmunu$ is independent of the hadron spin, and
the spin dependent part of $\wmunu$ is antisymmetric
in $\mu\nu$. We have already seen that the symmetric part of
$\ell_{\mu\nu}$ is independent of the lepton spin, and the
spin
dependent part of $\ell_{\mu\nu}$ is antisymmetric in $\mu\nu$.
Thus the combination $W^{\mu\nu}\ell_{\mu\nu}$
has no terms which have only the hadron spin, or only
the lepton spin; all terms contain either both or none. Thus
the structure functions $F_1$ and $F_2$ can be measured
using an unpolarised beam and target, but to measure the
structure functions $g_1$ and $g_2$ requires both a
polarised beam and a polarised target. There is no advantage
to doing an experiment with only a polarised beam or only a
polarised target. This is not true for target spins greater
than 1/2. For example, for a spin one target, there is
target polarisation
dependence in the symmetric part of $\wmunu$, which can
be measured
using a polarised target and an unpolarised beam.
\newsec{Scaling}

I have referred to scaling at several points so far. We are
now in a
position to understand what scaling means.
The tensor $\wmunu$ defined in
Eq.~\wmunudef\ is dimensionless, as are the structure
functions defined
in Eq.~\wmunuhalf.
The structure functions are dimensionless functions of the
Lorentz
invariant variables $p^2=M^2$, $p\cdot q$, and $q^2$. It is
conventional
to write them as functions of $x=Q^2/2 p\cdot q$ and $Q^2=-
q^2$, so they can be written as dimensionless functions
of
the dimensionless variables $x$ and $Q^2/M^2$. In elastic
scattering
there is a strong dependence on $Q^2/M^2$, and the elastic
form factors
fall off like a power of $Q^2/M^2$. It was thought that the
same
behaviour would persist for the deep inelastic structure
functions. The
scale $M$ is a typical hadronic scale, at which confinement
(and other
non-perturbative) effects
become important. Bjorken was the first to point out that if
the
constituents of the hadron were essentially free pointlike
objects at
high energies, then
the hadronic scale $M$ should be irrelevant, and the
structure functions
then only depend on $x$, and must be independent of
$Q^2$. This is
the famous prediction of scaling.

We now know that QCD is an asymptotically free theory, and
the
strong interaction coupling constant $\alpha_s$ becomes
small at
short distances. Thus at large $Q$, non-perturbative effects
(such as
the hadronic mass scale) are irrelevant, and QCD is
described by a
dimensionless coupling constant. Thus one recovers the
prediction of scaling in QCD, since there is no dimensionful
scale in
the problem. As is well known, scale invariance in a quantum
field
theory is broken because of quantum corrections. This introduces scaling
violations
because of
anomalous dimensions and the running of $\alpha_s$. Since
$\alpha_s$ is
small at high energies, these scaling violations can be
reliably
computed in QCD perturbation theory. Thus QCD predicts
scaling violations
(\ie\ $Q^2$ dependence)
for the structure functions which are calculable. We
will discuss this in more detail in Sec.~11.

\newsec{The Cross-Section for Spin-1/2 Targets}

The cross-section for spin-1/2 targets can now be obtained
by combining the expressions Eq.~\wmunuhalf\ for
$\wmunu$, Eq.~\lmunu\ for $\ell_{\mu\nu}$,
Eq.~\finalcross\ for the cross-section, and using the
identity
\eqn\xxyythree{
\epsilon^{\mu\nu\alpha\beta}\epsilon_{\mu\nu\lambda\sigma}=-
2\left(
g^\alpha_\lambda g^\beta_\sigma - g^\alpha_\sigma
g^\beta_\lambda\right),
}
\eqn\crosspol{\eqalign{
{d^2\sigma\over dx\, dy\, d\phi} = {e^4 M E\over 4 \pi^2
Q^4} \Bigm[
x y^2 F_1 &+ (1-y) F_2
+ y^2 g_1\left( 2 x\, {s_\ell \cdot s_h \over p \cdot q} + {q
\cdot s_\ell \over p \cdot q}{q \cdot s_h \over p \cdot q}
\right)\cr
&+ 2 x y^2 g_2 \left( {s_h \cdot s_\ell \over p \cdot
q}- {p \cdot s_\ell \over p \cdot q}{q \cdot s_h \over p
\cdot q} \right)\Bigm].\cr
}}
As mentioned earlier, the spin dependent terms involve both
$s_\ell$ and $s_h$.
Let us look at the cross-section for the two cases which are most
interesting
for current experiments, a longitudinally polarised lepton
beam incident on a target which is either longitudinally or
transversely polarised. For a longitudinally polarised
lepton beam, the polarisation is $s_\ell=\CH_\ell k$, where
$\CH_\ell=\pm$ is the lepton helicity. The lepton
polarisation terms in Eq.~\crosspol\ can be written as
\eqn\viiione{
{q \cdot s_\ell \over p \cdot q}=\CH_\ell\ {q \cdot k \over p
\cdot q}= -\CH_\ell\ x,\qquad
{p \cdot s_\ell \over p \cdot q}=\CH_\ell\ {p \cdot k \over p
\cdot q}={\CH_\ell\over y},\qquad
{s_h \cdot s_\ell \over p \cdot q}=\CH_\ell\ {s_h \cdot k
\over p \cdot q}.
}
\bigskip\vfill\break\eject
\leftline{{\it \underbar{Longitudinally Polarised Target}}}
\medskip
A target polarised along the incident beam direction has
$\vec s_h=M\CH_h \hat z$, where $\CH_h=\pm$ for a target
polarised parallel or antiparallel to the beam.
$s_h$ is dotted into either $q$ or $k$ both of which have
$0$ and $3$ components which are almost equal. Thus $s_h$
can be replaced by $-\CH_h\, p$ in the evaluation of the
cross-section in the deep inelastic limit.
(The additional minus sign occurs because of the relative
minus sign between the space and time components in a dot
product such as $k\cdot s$.) The result is
\eqn\longpol{
{d^2\sigma\over dx\, dy} = {e^4 M E\over 2 \pi Q^4} \left[
x y^2 F_1 + (1-y) F_2 - \CH_\ell\CH_h y(2-y)x g_1 +
\CO\left(M^2/Q^2\right)\right],
}
where the azimuthal angle $\phi$ has been integrated over,
since the cross-section is independent of $\phi$.
(The same expression can be used for HERA kinematics with the replacement
$2ME\rightarrow s$.)
Thus the polarisation asymmetry in the cross-section can be
used to measure the structure function $g_1$. The effect of
the structure function $g_2$ is suppressed relative to the
leading terms in Eq.~\longpol\ by an additional factor of
$M^2/Q^2$, and has been omitted.
\bigskip
\leftline{{\it \underbar{Transversely Polarised Target}}}
\medskip
The polarisation vector of a transversely polarised target
can be chosen to point along the $\hat x$ axis. In this case
\eqn\transpol{
{d^2\sigma\over dx\, dy\, d\phi} = {e^4 M E\over 2 \pi Q^4}
\left[
x y^2 F_1 + (1-y) F_2 + 2 x^2\,\CH_\ell\CH_h\   {M\over Q}
\sqrt{1-y}\,\cos\phi\,\left( g_1 y + 2 g_2\right)\right],
}
where I have used Eq.~\qcomps\ for the magnitude of the
transverse component of $q$. The structure functions $g_1$
and $g_2$ are equally important for a transversely polarised
target, and so an experiment with a transversely polarised
target can be used to determine $g_2$, once $g_1$ has been
measured using a longitudinally polarised target.
Note that the  spin-dependent piece of the cross-section for
a transversely polarised target is smaller than the spin-averaged
piece by a factor of $M/Q$, unlike in Eq.~\longpol\
where both terms are comparable in magnitude.
Thus a small misalignment of the target relative to the beam
direction can lead to a large contamination of the
cross-section asymmetry due to the fractional longitudinal
polarisation. It is therefore best to measure the $\cos\phi$
dependence of the polarisation asymmetry to measure $g_2$
using a transversely polarised target.

\newsec{Structure Function Inequalities}

There are some simple inequalities among structure functions
that follow from unitarity. The differential cross-section
Eq.~\longpol\ must be positive over the entire kinematic
region, and for any sign of $\CH_h$ and $\CH_\ell$. This
requires that
\eqn\ineqal{
x y^2 F_1 + (1-y) F_2 \ge x y (2 -y) \abs{g_1} \ge 0,
}
for all values of $x$ and $y$. In particular, the inequality
at $y=1$ implies that
\eqn\fgineq{
F_1\ge \abs{g_1} \ge 0.
}
The $n^{\rm th}$ moment of a function $f$ is defined by
\eqn\momdef{
M_n(f) \equiv \int_0^1 dx\ x^{n-1} f(x),
}
We will derive sum rules for the moments of $F_1$ and
$g_1$ in Sec.~9. There are useful inequalities amongst
the moments that must be satisfied because of the restriction,
Eq.~\fgineq. For example,
\eqn\gineq{\eqalign{
\abs{M_n(g_1)} &= \abs{\int_0^1 dx\ x^{n-1} g_1(x)}
\le \int_0^1 dx\ x^{n-1} \abs{g_1(x)}\cr&\le
\int_0^1 dx\ x^{n-1} F_1(x) = M_n(F_1),\cr
}}
using the triangle inequality
$\abs{a+b} \le \abs{a} + \abs{b}$.  Thus any moment of $g_1$
is bounded by the corresponding moment of $F_1$. There is
also the inequality
\eqn\fineq{
M_n(F_1) = \int_0^1 dx\ x^{n-1} F_1(x)
\ge \int_0^1 dx\ x^{n-1} x F_1(x) = M_{n+1}(F_1),
}
since $F_1\ge 0$ and $0\le x\le1$. Thus we find the sequence
of inequalities
\eqn\foneineq{
M_1(F_1)\ge \boxeqn{M_2(F_1)} \ge M_3(F_1) \ge
\boxeqn{M_4(F_1)}\ldots
}
\eqn\frelgineq{
M_1(F_1)\ge \boxeqn{\abs{M_1(g_1)}}\, , \quad
\boxeqn{M_2(F_1)}\ge \abs{M_2(g_1)},\quad
M_3(F_1)\ge \boxeqn{\abs{M_3(g_1)}}\, ,\quad
\ldots
}
The boxed quantities are the moments for which we will
derive QCD sum rules in Sec.~9. The second moment of
$F_1$, $M_2(F_1)$ bounds all the higher moments of $F_1$, as
well as all moments of $g_1$ for which there are sum rules,
except the first. This has an important experimental
consequence which will be studied in Sec.~10, since the
sum rule for the first moment of $g_1$ is the Ellis-Jaffe
sum rule.

\newsec{Helicity Amplitudes: The Interpretation of Structure
Functions}

The structure functions $F_1$, $F_2$, $g_1$, and $g_2$
defined in
Eq.~\wmunuhalf\ have simple physical interpretations. To
understand them better, it is useful to look at the
amplitude for forward Compton scattering off a hadron target
(see \fig\comptonfig{})
\eqn\tmunudef{
(T_{\mu\nu})_{\lambda'\lambda}=  i \int
d^4x\,e^{iq\cdot x}\bra{p,\lambda'} T\left(j_\mu(x)
j_\nu(x)\right)\ket{p,\lambda}.
}
\midinsert
\epsffile{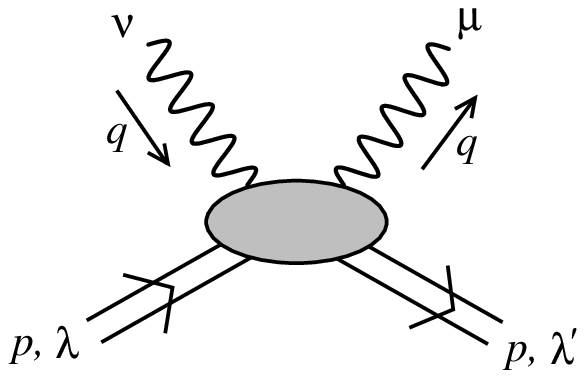}
\centerline{\tenrm FIGURE 3.}
\centerline{\tenrm The forward Compton scattering
amplitude.}
\endinsert\noindent
The tensor $T_{\mu\nu}(p,q,s_h)$ can be written in terms
$(T_{\mu\nu})_{\lambda'\lambda}$ by analogy with
Eq.~\wspinhalf,
\eqn\tspinhalf{
T_{\mu\nu}(p,q,s_h) = \sum_{\lambda,\lambda'}
\rho_{\lambda\lambda'}\,
T_{\mu\nu}(p,q)_{\lambda'\lambda} = \Tr \rho\, T_{\mu\nu}.
}
The tensor $T_{\mu\nu}(p,q,s_h)$ has the same symmetry
properties
as $\wmunu(p,q,s_h)$, Eqs.~\parityinv, \tinv, \herminv,
and the current conservation constraint
Eq.~\currcons. Thus $T_{\mu\nu}(p,q,s_h)$ can be expanded in
the same tensors as $\wmunu$. The structure
functions for $T_{\mu\nu}(p,q,s_h)$ will be denoted by a
tilde to distinguish them for the corresponding structure functions for
$\wmunu$:
\eqn\tstrfn{
\eqalign{T_{\mu\nu} &=  \tilde F_1 \left(-g_{\mu\nu} +
{q_\mu q_\nu\over q^2}\right)
+ {\tilde F_2\over p \cdot q} \left(p_\mu - {p\cdot q
\ q_\mu\over q^2}\right)
\left(p_\nu - {p\cdot q\ q_\nu\over q^2}\right)\cr
\noalign{\smallskip}&+ {i\, \tilde g_1\over p\cdot q}
\ \epsilon_{\mu\nu\lambda\sigma} q^\lambda s^\sigma +
{i\, \tilde g_2\over (p\cdot q)^2}
\ \epsilon_{\mu\nu\lambda\sigma} q^\lambda \left( p\cdot q
\, s^\sigma - s\cdot q\,p^\sigma\right)\ . \cr}
}
The photon crossing symmetry relation for $T_{\mu\nu}$
\eqn\tcrossinv{
T_{\mu\nu}(p,q,s_h) = T_{\nu\mu}(p,-q,s_h),
}
differs by a sign from Eq.~\crossinv\ for $\wmunu$, because
exchanging $j^\mu$ and $j^\nu$ inside a commutator produces
a minus sign, whereas exchanging them inside a time-ordered
product does not.
Eq.~\tcrossinv\ implies that $\tilde F_1$ is an even
function of $\omega$, and $\tilde F_2$, $\tilde g_1$, and
$\tilde g_2$ are odd functions of $\omega$,
\eqn\evenoddrln{\vcenter{\openup1\jot
\halign{$\hfil#$&${}#\hfil$
      &\qquad$\hfil#$&${}#\hfil$\cr
\tilde F_1(-\omega)&=\tilde F_1(\omega),&
\tilde F_2(-\omega)&=-\tilde F_2(\omega),\cr
\tilde g_1(-\omega)&=-\tilde g_1(\omega),&
\tilde g_2(-\omega)&=-\tilde g_2(\omega).\cr
}}}
The Compton
amplitude for photon-antinucleon scattering can be obtained
from the photon-nucleon scattering amplitude by fermion crossing
symmetry, \ie\ by replacing an outgoing hadron with momentum
$p$ and spin $s_h$ by an incoming antihadron
with momentum $-p$ and spin $s_h$.
Thus the  Compton amplitude for antinucleon scattering is
$T_{\mu\nu}(-p,q,s_h)$. This implies
\eqn\fcrossrln{\vcenter{\openup1\jot
\halign{$\hfil#$&${}#\hfil$
      &\qquad$\hfil#$&${}#\hfil$\cr
\tilde F_1^{\rm antinucleon}(\omega)&=\tilde F_1^{\rm nucleon}(-\omega),&
\tilde F_2^{\rm antinucleon}(\omega)&=-\tilde F_2^{\rm nucleon}(-\omega),\cr
\tilde g_1^{\rm antinucleon}(\omega)&=-\tilde g_1^{\rm nucleon}(-\omega),&
\tilde g_2^{\rm antinucleon}(\omega)&=-\tilde g_2^{\rm nucleon}(-\omega).\cr
}}}
using Eq.~\tstrfn\ and noting that $\omega\rightarrow-\omega$ if
$p\rightarrow -p$.  Eqs.~\evenoddrln\ and \fcrossrln\ imply that the
antinucleon structure functions are equal to  the corresponding nucleon
structure
functions,
\eqn\equalrln{\vcenter{\openup1\jot
\halign{$\hfil#$&${}#\hfil$
      &\qquad$\hfil#$&${}#\hfil$\cr
\tilde F_1^{\rm antinucleon}(\omega)&=\tilde F_1^{\rm nucleon}(\omega),&
\tilde F_2^{\rm antinucleon}(\omega)&=\tilde F_2^{\rm nucleon}(\omega),\cr
\tilde g_1^{\rm antinucleon}(\omega)&=\tilde g_1^{\rm nucleon}(\omega),&
\tilde g_2^{\rm antinucleon}(\omega)&=\tilde g_2^{\rm nucleon}(\omega).\cr
}}}
This relation could have been obtained trivially by applying charge
conjugation to $T_{\mu\nu}$, since it converts the nucleon to an
antinucleon. There are no additional signs because the product of two
electromagnetic currents is even under charge conjugation.

The optical theorem implies that twice the imaginary part of
the forward scattering amplitude is the total cross-section.
There are factors of $i$ in the expansion of $T_{\mu\nu}$
and $\wmunu$, so what the optical theorem implies in our
case is that twice the imaginary part of the $T_{\mu\nu}$
structure functions is equal to corresponding $\wmunu$
structure functions times $4 \pi$. The extra factor of $4\pi$
is due to the extra factor of $1/4\pi$ in the definition of
$\wmunu$ relative to $T_{\mu\nu}$.
The analytic structure of the amplitudes $\tilde
F_1$, \etc\ is shown in \fig\figanalytic{}\
as a function of $\omega=1/x$.
\midinsert
\epsffile{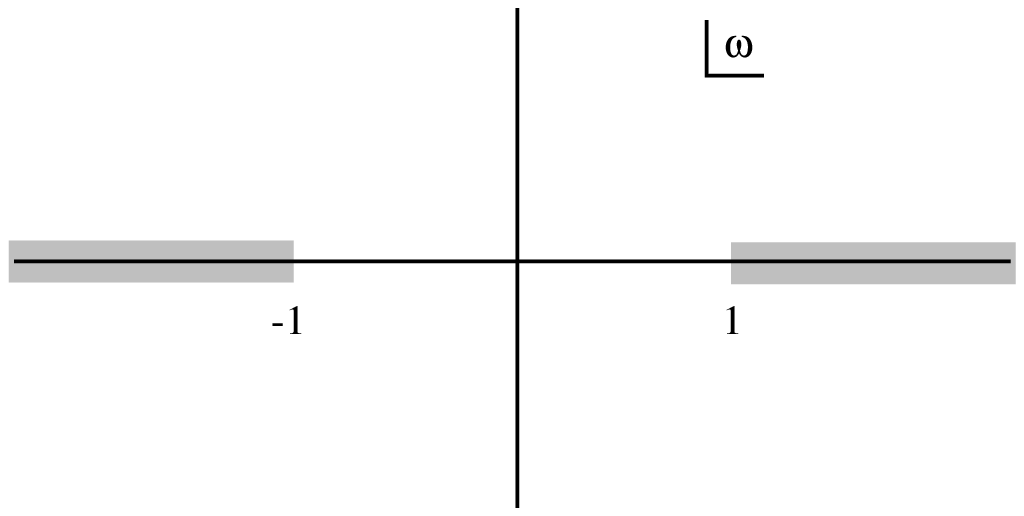}
\centerline{\tenrm FIGURE 4.}
{\tenrm \noindent The analytic
structure of $\scriptstyle T_{\mu\nu}$ in the complex $\scriptstyle \omega$
plane. The
discontinuity across the cuts $\scriptstyle 1\le \abs{\omega}\le\infty$
is related to $\scriptstyle \wmunu$.}
\endinsert\noindent
There are cuts in the physical region
$1\le\abs{\omega}\le\infty$. The optical theorem implies that
\eqn\optthm{\vcenter{\openup1\jot
\halign{$\hfil#$&${}#\hfil$
      &\qquad$\hfil#$&${}#\hfil$\cr
\Im \tilde F_1(\omega + i \epsilon) &= 2 \pi F_1(\omega),&
\Im \tilde F_2(\omega + i \epsilon) &= 2 \pi F_2(\omega),\cr
\Im \tilde g_1(\omega + i \epsilon) &= 2 \pi g_1(\omega),&
\Im \tilde g_2(\omega + i \epsilon) &= 2 \pi g_2(\omega).\cr
}}}
The discontinuity of $T_{\mu\nu}$ across the
right hand cut gives $\wmunu$ for nucleon targets.
The antinucleon structure functions $F_1$, $F_2$, $g_1$ and $g_2$ are
equal to the corresponding nucleon structure functions, because of
Eq.~\equalrln, a result which also follows trivially from charge conjugation.
We also see from Eq.~\fcrossrln\
that the discontinuity across the left hand cut in the complex $\omega$
plane is the physical antinucleon structure function.

%

The forward Compton amplitude can be expanded in terms of
helicity amplitudes. Pick the $\hat z$ axis to be along the
direction of the virtual photon, and choose this axis
to quantise the angular momentum of the target and photon.
The initial and photon spin components along $\hat z$ will
be called $h$ and $h'$, and the initial and final target
spin components will be called $H$ and $H'$. The helicity
amplitude for forward Compton scattering (see \fig\hampfig{}):
\midinsert
\epsffile{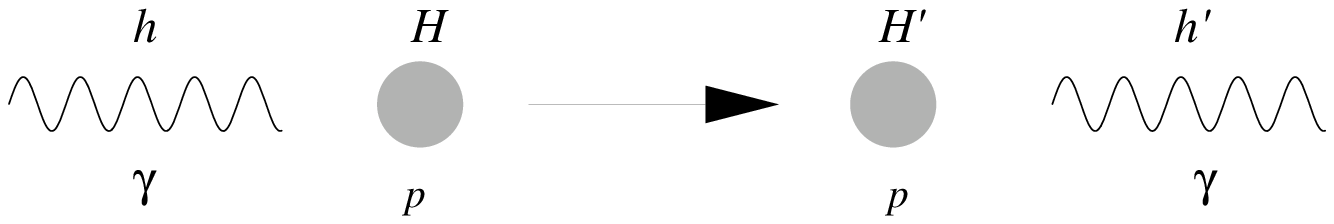}
\centerline{\tenrm FIGURE 5.}
{\tenrm\noindent The incoming photon scatters off the target proton which is at
rest. The helicity amplitude for this process is denoted by $\scriptstyle
\CA(h,H;h',H')$.}
\endinsert
\eqn\xone{
\gamma_h + {\rm target}_H \rightarrow \gamma_{h'} + {\rm
target}_{H'}
}
will be denoted by $\CA(h,H;h',H')$.
Angular momentum conservation along the $\hat z$ axis implies that
\eqn\angmom{
h+H=h'+H'.
}
Reflecting in the plane of the scattering does not change
any of the momenta, but reverses all the spins. Thus parity
invariance implies that
\eqn\hapinv{
\CA(h,H;h',H') = \CA(-h,-H;-h',-H').
}
Time reversal exchanges initial and final states, and
reverses the directions of momenta and spin. Since the
helicity of the photons and the target is defined as the
component of spin along the direction of motion of the
photon, helicity does not change sign under time reversal.
Thus time reversal invariance implies that
\eqn\hatinv{
\CA(h,H;h',H')=\CA(h',H';h,H).
}
The virtual photon has momentum $q=(q^0,0,0,q^3)$, with
three possible polarisations,
\eqn\photonpol{\eqalign{
\epsilon_+^\mu &= -{1\over\sqrt{2}}(0,1,i,0),\quad h=+1,\cr
\epsilon_-^\mu &= {1\over\sqrt{2}}(0,1,-i,0),\quad h=-1,\cr
\epsilon_0^\mu &= {1\over Q} (q^3,0,0,q^0), \quad\ h=0,\cr
}}
where $\epsilon_{\pm}$ are spacelike, and $\epsilon_0$ is
timelike.
The fourth polarisation proportional to $q^\mu$ does not
contribute since the electromagnetic current is conserved.
Denoting $H=1/2$ by $\uparrow$, and $H=-1/2$ by $\downarrow$
and using Eqs.~\angmom--\hatinv, we see that for a spin-1/2
target there are only four independent helicity amplitudes
which can be chosen to be
\eqn\indepamp{
\CA(+,\uparrow;+,\uparrow),\quad
\CA(+,\downarrow;+,\downarrow), \quad
\CA(0,\uparrow;0,\uparrow), \quad
\CA(+,\downarrow;0,\uparrow).
}
There are four independent structure functions for spin-1/2
targets because there are four independent helicity
amplitudes. The helicity amplitudes can be computed in terms
of $T_{\mu\nu}$ using
\eqn\xtwo{
\CA(h,H;h',H')=\epsilon_{h'}^{\mu*}\,\epsilon_{h}^\nu
\ T_{\mu\nu}(s),
}
with
\eqn\xthree{
\vec s = M\ \bar u(H')\, \vec \sigma\, u(H).
}
For spin-1/2 targets, $u(\uparrow) = \pmatrix{1\cr0\cr}$ and
$u(\downarrow) = \pmatrix{0\cr1\cr}$. The computation of Eq.~\xtwo\ is
simple for spin-1/2 targets in the deep inelastic limit.
Using Eq.~\tstrfn\ and neglecting terms of order $M/Q$, we
find
\eqn\hasf{\vcenter{\openup1\jot
\halign{$\hfil#$&${}#\hfil$
      &\qquad$\hfil#$&${}#\hfil$\cr
\CA(+,\uparrow;+,\uparrow)&= \tilde F_1-\tilde g_1,&
\CA(+,\downarrow;+,\downarrow)&=\tilde F_1+\tilde g_1,\cr
\noalign{\medskip}
\CA(0,\uparrow;0,\uparrow)&={\tilde F_2/ 2x \tilde F_1}
\equiv \tilde F_L , & \CA(+,\downarrow;0,\uparrow)&={\left(2\sqrt 2 x
M/Q\right)}(\tilde
g_1+\tilde g_2).
\cr}}}
For example, the coefficient of $\tilde F_1$ in Eq.~\tstrfn\
is $-g_{\mu\nu}$. Thus the coefficient of $\tilde F_1$ in
Eq.~\hasf\ is $+1$ for $\CA(+,\uparrow;+,\uparrow)$ and
$\CA(+,\downarrow;+,\downarrow)$ because $\epsilon_+$ is
spacelike, is $-1$ for $\CA(0,\uparrow;0,\uparrow)$
because $\epsilon_0$
is timelike, and is zero for $\CA(+,\downarrow;0,\uparrow)$
because $\epsilon_+$ and $\epsilon_0$ are orthogonal.
Taking imaginary parts of the left hand side gives identical
formul\ae\ with $\tilde F\rightarrow 2\pi F$, \etc\ Thus we
see that
\eqn\xthree{
2 \pi F_1 = {1\over 2}\Im
 \left[\CA(+,\uparrow;+,\uparrow)+\CA(+,\downarrow;+,\downarrow)\right],
}
is the cross-section for scattering a transverse
photon off an unpolarised target,
\eqn\xfour{
2 \pi g_1 = -{1\over 2}\Im
\left[\CA(+,\uparrow;+,\uparrow)-
\CA(+,\downarrow;+,\downarrow)\right]
}
is the spin asymmetry in the scattering cross-section for a
transverse photon, and
\eqn\xfive{
2 \pi F_L = \Im \CA(0,\uparrow;0,\uparrow) = \Im
\CA(0,\downarrow;0,\downarrow)
}
is the cross-section for scattering longitudinally polarised
photons off an unpolarised target. $F_L$ is called the
longitudinal structure function. The physical interpretation
of $g_2$ is obscure, the combination $g_1+g_2$ is
proportional to the ``single helicity flip''
amplitude $\CA(+,\downarrow;0,\uparrow)$. Note that the
single helicity flip amplitude is suppressed by a factor of
$M/Q$ relative to the diagonal helicity amplitudes.

The helicity amplitude formalism is useful in discussing targets with spin
greater than 1/2. It provides a simple way of counting the number of
independent structure functions and understanding their physical significance.

\newsec{The Parton Model}

The parton model provides a simple physical picture which helps
in interpreting the structure functions of deep inelastic
scattering. The hadron target is considered to be made up of a
number of free on-shell partons (quarks and gluons) in the
deep inelastic limit. The scattering cross-section can then
be computed in terms of the incoherent scattering of free
quarks and gluons.
The hadronic tensor $W_{\mu\nu}$ can be computed in terms of
parton distribution functions. For example, the distribution function
$u_\uparrow(\xi)$ is the probability to find an up quark in
the hadron with polarisation parallel to that of the hadron,
and momentum $\xi p$. It is conventional to go to an
infinite momentum frame when discussing the parton model,
and to claim that partons are non-interacting in this
infinite momentum frame. This is incorrect. Partons can be
treated as free because of asymptotic freedom, not because
of the choice of a particular Lorentz frame. At high
energies, the QCD coupling constant becomes weak, and so one
can treat parton interactions using perturbation theory. The
main reason for going to the infinite momentum frame is to
neglect target mass effects. A parton with momentum fraction
$\xi$ has invariant mass $\xi^2 M^2$, which is not zero, and
depends on $\xi$. The infinite momentum frame hides this
embarrassment, which is why it is used to discuss the parton
model. Instead of using the infinite momentum frame, I will work in the rest
frame of the target,
and consistently neglect all target mass effects.

To lowest order in the strong interactions, only quarks
scatter off the virtual photon, since gluons do not have
electric charge. The contribution to $W_{\mu\nu}$ from a
single polarised quark can be computed from the Feynman
graph of \fig\quarkparton{}. The
computation of the scattering amplitude can be obtained from
\midinsert
\epsffile{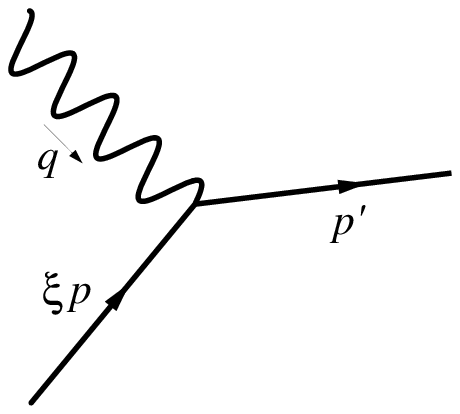}
\centerline{\tenrm FIGURE 6.}
\centerline{\tenrm The contribution to $\scriptstyle W_{\mu\nu}$
from a single quark with momentum fraction $\scriptstyle \xi$.}
\endinsert\noindent
$\ell_{\mu\nu}$, Eq.~\lmunu, by the replacements $k\rightarrow
\xi p$, $k'\rightarrow p'$, $s_\ell\rightarrow s$ and
$q \rightarrow -q $, and multiplying by the
square of the quark charge $\CQ$.
Comparing Eq.~\lmunu\ and Eq.~\wmunutwo, we see
that $W_{\mu\nu}$ has an additional factor of integration
over the final particle phase space. Thus we find
\eqn\wmunuquark{\eqalign{
W_{\mu\nu}={1\over 4 \pi}\CQ^2\int& {d^3 p'\over (2\pi)^3 2
E_{p'}}\ {1\over\xi}\
 (2\pi)^4 \delta^4(\xi p + q - p')\cr &\quad\times 2 \left[ \xi p^\mu
p'^\nu +
\xi p^\nu p'^\mu - g^{\mu\nu} \xi p\cdot p' + i
\epsilon^{\mu\nu\alpha\beta} q_{\alpha} s_{\beta}\right].\cr
}}
The only subtlety is the factor of $1/\xi$. The proton
states in $\wmunu$ are normalised to $2p^0$, whereas the
parton states in $\wmunuquark$ are normalised to $2p^0\xi$
in the standard relativistic normalisation convention. This
affects the particle flux in the computation of the cross-section.
The extra factor of $1/\xi$ converts the parton
flux to the correct normalisation for the proton flux.
The integral in Eq.~\wmunuquark\ is easily done using the identity
\eqn\xione{
\int {d^3 p'\over (2\pi)^3 2 E_{p'}} = \int {d^4 p'\over
(2\pi)^4}
\ (2\pi)\, \delta\left(\left(\xi p + q - p'\right)^2\right) = \int {d^4 p'\over
(2\pi)^4}
\ {(2\pi)\over 2 p \cdot q}\ \delta\left(\xi  + {q^2\over 2 p \cdot
q}\right).
}
Since
the partons are massless, $s_\beta = h \xi p_\beta$, where
$h$ is the helicity of the parton, so that
\eqn\xitwo{\eqalign{
W_{\mu\nu}&={\CQ^2\over 2\xi\  p\cdot q}\left[ \xi p^\mu p'^\nu +
\xi p^\nu p'^\mu - g^{\mu\nu} \xi p\cdot p' + i h \xi
\,\epsilon^{\mu\nu\alpha\beta} q_{\alpha}
p_{\beta}\right]\delta(\xi-x),\cr
&= {\CQ^2\over 2\xi\ p\cdot q}\left[ 2\xi^2 p^\mu p^\nu
- g^{\mu\nu} \xi p\cdot q + i h  \xi
\,\epsilon^{\mu\nu\alpha\beta} q_{\alpha}
p_{\beta}\right]\delta(\xi-x),
}}
where $p'$ has been replaced by $\xi p + q$, and the $q^\mu$
and $q^\nu$ terms have been dropped.
To compare with $\wmunu$, we need Eq.~\wmunuhalf\
neglecting target mass effects. Thus the target spin $s_h$
in Eq.~\wmunuhalf\ can be replaced by $\CH p$, where $\CH$
is the helicity of the target, and $p$ is the target momentum.
Comparing with Eq.~\wmunuhalf, we see that the
contribution
of quarks to the
structure functions is
\eqn\xithree{
F_1 = {\CQ^2\over2} \delta(\xi-x),\quad F_2 = \CQ^2\xi \, \delta(\xi-
x),\quad
g_1 = {\CQ^2\over2} h\CH\,\delta(\xi-x),\quad g_2=0.
}
The antiquark contribution can be computed similarly, and is
identical to the quark contribution.
The total structure function is obtained by integrating
with the quark distribution functions, $q_+(\xi)$, the
probability to find a quark in the nucleon with momentum
fraction $\xi$ and helicity equal to that of the nucleon
($h\CH=+$), and $q_-(\xi)$, the probability to find a quark
in the nucleon with momentum fraction $\xi$, and helicity
opposite to that of the nucleon ($h\CH=-$), and the antiquark
distribution functions $\bar q_{\pm}(\xi)$:
\eqn\partonsf{\eqalign{
F_1(x,Q^2) &= \sum_i {\CQ_i^2\over2}\,  \left( q_+(x)+
q_-(x)+\bar q_+(x)+ \bar q_-(x) \right),\cr
F_2(x,Q^2) &= \sum_i \CQ_i^2\ x\, \left(q_+(x)+
q_-(x) + \bar q_+(x)+ \bar q_-(x)\right),\cr
g_1(x,Q^2) &= \sum_i {\CQ_i^2\over2}\, \left(q_+(x) -
q_-(x)+\bar q_+(x) - \bar q_-(x)\right),\cr
g_2(x,Q^2) &= 0,
}}
where the sum is over all quark flavours which can be considered
light compared with $Q^2$.
The most important feature of this result is that the
structure functions depend only on $x$, and are independent
of $Q^2$. We also see that the structure functions satisfy
the Callan-Gross relation,
\eqn\xifour{
F_2(x) = 2 x F_1(x) \quad\Rightarrow \quad F_L(x)=0.
}
The Callan-Gross relation will get corrections at
order $\alpha_s$ in QCD. The structure function $g_2$
vanishes identically; there is no simple interpretation for
$g_2$ in the parton model.

We see from Eq.~\partonsf\ that $F_1$ is the probability to
find a quark in the hadron
with momentum fraction $x$, and $g_1$ is the difference in
probabilities to find a quark in the hadron with momentum
fraction $x$ with spin parallel and antiparallel to the
hadron. Both quantities weight each quark flavour by the
square of the electric charge.
The combination $F_1(x)-g_1(x)$ depends only on $q_-(x)$.
The reason is that $F_1(x)-g_1(x)$ is proportional to the
imaginary part of
the helicity amplitude $\CA(+,\uparrow;+,\uparrow)$. There
are two possible diagrams which contribute to the Compton
scattering amplitude, one where the
quark absorbs the initial photon, and then emits the
final photon, and the crossed diagram where the final
photon is emitted before the initial photon is absorbed. The
crossed diagram has no imaginary part since the energy of
the initial and final photons is positive, and $q^2<0$. The intermediate
state when a quark with $J_z=m$ absorbs a helicity $+$
photon is a virtual quark with $J_z=h+1$. A virtual quark
only has $J_z=\pm 1/2$, so only $J_z=-1/2$ quarks can
absorb a helicity $+$ photon. This implies that only $q_-(x)$
contributes to $F_1(x)-g_1(x)$, and similarly, only $q_+(x)$
contributes to $F_1(x)+g_1(x)$.

\bigskip
\leftline{{\it \underbar{Isospin Symmetry and the Gottfried Sum Rule}}}
\medskip
It is conventional to use the symbol $q_{\pm}(x)$ to refer
to the parton distributions in a proton target. The
distributions in a neutron target are related to those in a
proton target by isospin invariance. Making a $180^\circ$
rotation in isospin space exchanges $p\leftrightarrow n$ and
$u\leftrightarrow d$. Thus one finds
\eqn\isospin{\vcenter{\openup1\jot
\halign{$\hfil#$&${}#\hfil$
      &\qquad$\hfil#$&${}#\hfil$\cr
u_{\pm,\rm\ proton}(x) &= d_{\pm,\rm\ neutron}(x),&
\bar u_{\pm,\rm\ proton}(x) = \bar d_{\pm,\rm\ neutron}(x),\cr
d_{\pm,\rm\ proton}(x) &= u_{\pm,\rm \ neutron}(x),&
\bar d_{\pm,\rm\ proton}(x) = \bar u_{\pm,\rm\ neutron}(x),\cr
s_{\pm,\rm\ proton}(x) &= s_{\pm,\rm \ neutron}(x),&
\bar s_{\pm,\rm\ proton}(x) = \bar s_{\pm,\rm\ neutron}(x),\cr
}}}
Isospin invariance is broken by light quark masses, and by
electromagnetism.
The light quark mass effects are of order $m_q/\lqcd$, and the
electromagnetic corrections are of order $\alpha/4\pi$, so one expects
isospin violation at the few percent level. There is no large isospin
breaking effect proportional to $m_u/m_d$; the relevant quantity
in QCD is the ratio of a light quark mass to the hadronic scale $\lqcd$.
There is no reason whatsoever for the ``quark sea to be
isospin symmetric,'' \ie\ for
\eqn\wrongiso{
\bar u_{\rm proton}(x)=\bar d_{\rm proton}(x),
}
because the proton state is not isospin symmetric. One
simple way to see this is to note that the production of
$u\bar u$ pairs in the proton must be different from the production
of $d \bar
d$ pairs in the proton, since there are two ``net'' $u$
quarks in the proton, and only one ``net'' $d$ quarks, and
any $u$ quarks produced must be antisymmetric with respect
to the existing quarks.

There is a (false) sum rule for the first moment of $F_1^{ep}-F_1^{en}$, the
difference in $F_1$ between protons and neutrons. This is the
Gottfried sum rule which can be derived as follows:
\eqn\gott{\eqalign{
\int_0^1 F_1^{ep}(x)-F_1^{en}(x) &= \int_0^1
\left[ \frac49 \left[u(x)+\bar u(x)\right] + \frac19
\left[d(x) +\bar d(x) \right] +\frac 19 \left[s(x)+\bar
s(x)\right]\right]_{\rm proton}\cr
&\quad-\left[ \frac49 \left[u(x)+\bar u(x)\right] + \frac19
\left[d(x) +\bar d(x) \right] +\frac 19 \left[s(x)+\bar
s(x)\right]\right]_{\rm neutron}\cr
&= \frac13 \int_0^1
\left[ u(x) - d(x) + \bar u(x)- \bar d(x)\right]_{\rm
proton}\cr
&=\frac13 \quad({\sl Warning:\ This\ is\ false}).
}}
The second line follows from Eqs.~\isospin, and the last line
follows from using Eq.~\wrongiso\ to write
\eqn\xiione{
\frac13 \int_0^1
\left[ u(x) - d(x) + \bar u(x)- \bar d(x)\right]_{\rm
proton}= \frac13 \int_0^1
\left[ u(x) - d(x) \right] = \frac13,
}
but since Eq.~\wrongiso\ is incorrect, so is the Gottfried
sum rule.
There is another reason why the Gottfried sum rule is
incorrect; it is an attempt to derive a sum rule for the
``wrong'' moment of $F_1$ as we will discuss later. I
repeat, {\sl the Gottfried sum rule is false in QCD.}

\newsec{The Operator Product Expansion and Sum Rules}

The parton model provides an interpretation for the deep
inelastic structure functions.
One can instead derive sum rules for
the structure functions directly from QCD using the operator
product expansion.
These sum rules are extremely important because they do not
depend on any model of hadronic structure, and provide a direct test
of QCD. They can be derived using only some very general results
from quantum field theory.
The
starting point in the derivation is the time-ordered product of two currents
\eqn\tmunu{
t_{\mu\nu} \equiv i \int d^4x\,e^{iq\cdot x} T\left(j_\mu(x)
j_\nu(0)\right).
}
The nucleon  matrix element of $t_{\mu\nu}$ gives the
familiar Compton amplitude,
\eqn\xiiione{
(T_{\mu\nu})_{\lambda'\lambda}
=\bra{p,\lambda'}t_{\mu\nu}\ket{p,\lambda}.
}
The analytic structure of $T_{\mu\nu}$ and its relation to
$\wmunu$ was discussed in Sec.~7.

The key idea that permits the computation of $T_{\mu\nu}$ in
QCD is the operator product expansion. Consider the product
of two local operators,
\eqn\xiiitwo{
\CO_a(x)\ \CO_b(0).
}
In the limit that $x\rightarrow 0$, the operators are at
practically the same point. In this limit,
the operator product can be written as an expansion in local
operators,
\eqn\opeeg{
\lim_{x\rightarrow0}\CO_a(x)\ \CO_b(0) = \sum_k c_{abk}(x)\ \CO_k(0).
}
The coefficient functions depend on the separation $x$.  The
left hand side is completely equivalent to the right hand
side as long as one does not probe the operator product on
distance scales which are small compared with the separation
$x$. Thus one can replace the product
$\CO_a(x) \CO_b(0)$ in the computation of matrix elements by
the expansion Eq.~\opeeg\ where the
coefficients $c_{abk}(x)$ {\sl are independent of the matrix
elements, provided that the external states have momentum
components which are small compared with the separation
$x$.} In QCD, the coupling constant is small at short
distances because of asymptotic freedom. Thus the
coefficient functions can be computed in perturbation
theory, since all  non-perturbative effects occur at scales
which are much larger than $x$, and thus do not affect the
computation of the coefficient functions.

The momentum space version of the operator product expansion is for
the product
\eqn\xiiithree{
\int d^4x\, e^{i q \cdot x}\ \CO_a(x)\ \CO_b(0).
}
In the limit that
$q\rightarrow\infty$, the Fourier
transform in Eq.~\tmunu\ forces $x\rightarrow 0$, and again the operator
product can be expanded in terms of local operators with coefficient functions
that depend on $q$,
\eqn\xiiifour{
\lim_{q\rightarrow\infty}\int d^4x\, e^{i q \cdot x}\ \CO_a(x)\ \CO_b(0)
 = \sum_k c_{abk}(q)\ \CO_k(0).
}
 This expansion is valid for all matrix elements, provided $q$ is much larger
than the characteristic momentum in any of the external states.

We will use the Fourier transform version of the operator product expansion,
Eq.~\xiiifour. The product of two electromagnetic currents in Eq.~\tmunu\ can
be expanded in terms of a sum of local operators multiplied by coefficients
which are functions of $q$. This expansion will be valid for matrix elements in
the target, Eq.~\xiiione, provided that $q$ is much larger than the typical
hadronic mass scale $\lqcd$.
The local operators in the operator product expansion for QCD are
quark and
gluon operators with arbitrary dimension $d$ and spin $n$.
An operator
with spin $n$ and dimension $d$ can be written as
\eqn\xiiifive{
\CO^{{\mu_1}\ldots {\mu_n}}_{d,n},
}
where $\CO_{d,n}$ is symmetric and traceless in
$\mu_1\ldots\mu_n$. The matrix
element of $\CO_{d,n}$ in the hadron target is proportional to
\eqn\xiiisix{
M^{d-n-2}\ \CS \left[ p^{{\mu_1}}\ldots p^{\mu_n}\right],
}
for a vector operator, and to
\eqn\xiiiseven{
M^{d-n-2}\ \CS\left[s^{{\mu_1}}p^{{\mu_2}}\ldots
p^{\mu_n}\right],
}
for an axial operator. $\CS$ acts on a tensor to project out
the
completely symmetric traceless component. The power of $M$
follows from
dimensional analysis, since a hadron state with the
conventional
relativistic normalisation has dimension minus one. The
coefficient functions
in the operator product expansion are functions only of $q$.
Thus the
free indices on the operator $\CO$ must be either $\mu,\nu$,
or be
contracted with $q^\alpha$. Every index on $\CO$ contracted
with
$q^\alpha$ produces a factor of $p\cdot q$ (or $s\cdot q$)
which is of
order $Q^2/M$ in the
deep inelastic limit. An index $\mu$ or $\nu$ is contracted
with
$\ell_{\mu\nu}$, and produces a factor of $p\cdot k$ or
$p\cdot k'$ (or
$s\cdot k$ or $s\cdot k'$), both of which are also of order
$Q^2/M$ in the
deep inelastic limit. In addition, since $t_{\mu\nu}$ has
dimension two,
the coefficient of $\CO$ must have dimension $Q^{2-d}$ in the
operator
product expansion.
Thus the contribution of any operator $\CO$ to
$W_{\mu\nu}\ell^{\mu\nu}$ is of order
\eqn\xiiieight{
\eqalign{
c_{{\mu_1}\ldots {\mu_n}} \CO^{{\mu_1}\ldots {\mu_n}}_{d,n}
&\rightarrow {q_{\mu_1}\over Q}\ldots {q_{\mu_n}\over Q}
\ Q^{2-d} \vev{\CO^{{\mu_1}\ldots {\mu_n}}_d},\cr
\noalign{\smallskip}&\rightarrow {q_{\mu_1}\over Q}\ldots {q_{\mu_n}\over Q}
\ Q^{2-d} M^{d-n-2}\  p^{\mu_1}\ldots
p^{\mu_n},\cr\noalign{\smallskip}&\rightarrow {(p\cdot q)^n \over Q^n}\
Q^{2-d}\ M^{d-n-2},\cr\noalign{\smallskip}&\rightarrow
\omega^n \left({Q\over M}\right)^{2+n-d}=
\omega^n \left({Q\over M}\right)^{2-t},\cr
}}
where the twist $t$ is defined as
\eqn\xiiinine{
 {\rm twist}=t  = d -n ={\rm dimension} - {\rm spin}.
}
The most important operators in the operator product expansion are those with
the lowest possible twist.  Twist two operators contribute a finite amount to
the structure functions in the deep inelastic limit, twist three contributions
are suppressed by $M/Q$, \etc\
The fundamental fields in QCD are quark and gluon fields, so the operators in
the operator product expansion can be written in terms of quark fields $\psi$,
the gluon field strength $G_{\mu\nu}$ and the covariant derivative $D^\mu$. We
can make a table listing the basic objects, with their dimension and twist:
\medskip
\centerline{\vbox{\offinterlineskip
\halign{\strut \hfil$\quad#\quad$\hfil&\vrule#&&\strut
\hfil$\quad#\quad$\hfil\cr
  & & \psi & G_{\mu\nu} & D^\mu\cr
\omit&height2pt&\omit&\omit&\omit\cr
\noalign{\hrule}
\omit&height2pt&\omit&\omit&\omit\cr
d & & 3/2 & 2 & 1\cr
\omit&height4pt&\omit&\omit&\omit\cr
s & & 1/2 & 1 & 1\cr
\omit&height4pt&\omit&\omit&\omit\cr
t & & 1 & 1 & 0 \cr
}}}
\medskip\noindent
Any gauge invariant operator must contain at least two quark fields, or two
gluon field strength tensors. Thus the lowest possible twist is two. A twist
two operator has either two $\psi$'s or two $G_{\mu\nu}$'s and an arbitrary
number of covariant derivatives. The indices of the covariant derivatives are
not contracted, because an operator such as $D^2$ has twist two, whereas
$D^\alpha D^\beta$ has twist zero.

The first step in doing an operator product expansion is to determine all the
linearly independent operators that can occur.
We have just seen that the leading operators are twist two quark and gluon
operators. The Lorentz structure of the quark operators must be either $\bar
\psi \gamma^\mu \psi$ or $\bar \psi \gamma^\mu \gamma_5 \psi$ in the limit that
light quark masses can be neglected, because the operator $j^\mu j^\nu$ does
not change chirality. The conventional basis for twist two quark operators is:
\eqn\ov{
O^{\mu_1\ldots \mu_n}_{V,a}= {1\over 2} \left( {i\over
2}\right)^{n-1} S \left\{\bar{\psi_a} \,\gamma^{\mu_1} \darr{D}^{\mu_2} \ldots
\darr{D}^{\mu_n}\psi_a\right\}\ ,
}
\eqn\oa{
O^{\mu_1\ldots \mu_n}_{A,a} = {1\over 2} \left( {i\over 2}\right)^{n-1} S
\left\{\bar{\psi_a} \,\gamma^{\mu_1} \darr{D}^{\mu_2} \ldots
\darr{D}^{\mu_n} \gamma_5 \psi_a\right\}\ ,
}
where the index $a=(u,d,s)$ runs over the light quark flavours.
The twist two gluon operators are more complicated, and I will give one such
operator tower:
\eqn\gluonopone{
O_{g,V}^{\mu_1\ldots\mu_n} = -{1\over 2}\left({i\over 2}\right)^{n-2}
S\left\{ G_a^{\mu_1\alpha}\darr{D}^{\mu_2} \ldots \darr{D}^{\mu_{n-1}}
G_{a\alpha}{}^{\mu_n}\right\}\ .}
In these lectures, I will only compute the operator product expansion to lowest
order, so I do not need the gluon operators.

The second step in doing an operator product expansion is to determine the
coefficient functions of the operators. The best way to do this to to evaluate
enough on-shell matrix elements to determine all the coefficients. Since we
have argued that the coefficients can be computed reliably in QCD perturbation
theory, we will evaluate the coefficients by taking matrix elements in on-shell
quark and gluon states. I will only illustrate the computation of the
coefficients to lowest non-trivial order (\ie\ $(\alpha_s)^0$) in this lecture.
The generic term in the operator product expansion can be written as
\eqn\xiiieleven{
j j \sim c_q \CO_q + c_g \CO_g,
}
where $q$ and $g$ refer to quark and gluon operators. Taking the matrix element
of both sides in a free quark state, gives
\eqn\xiiitwelve{
\bra{q} j j \ket{q} \sim c_q \bra{q} \CO_q \ket {q} + c_g \bra{q} \CO_g
\ket{q}.
}
The electromagnetic current is a quark operator. Thus the left hand side is of
order $(\alpha_s)^0$. The matrix element
$\bra{q} \CO_q \ket {q}$ is also of order $(\alpha_s)^0$, whereas the matrix
element $\bra{q} \CO_g \ket{q}$ is of order
$(\alpha_s)^1$ since there are at least two gluons in $\CO_g$, each of which
contributes a factor of the QCD coupling constant $g$ to the matrix element.
Thus one can determine $c_q$ to leading order by taking the quark matrix
element of both sides of the operator product expansion, neglecting the gluon
operators. I will work for simplicity in a theory with a single quark flavour
with charge one, and generalise the result to an arbitrary number of flavours
with arbitrary charges at the end.

The quark matrix element of the left hand side of the
operator product expansion, Eq.~\xiiieleven, is
given by the Feynman graphs in  \fig\opefig{},
\midinsert
\epsffile{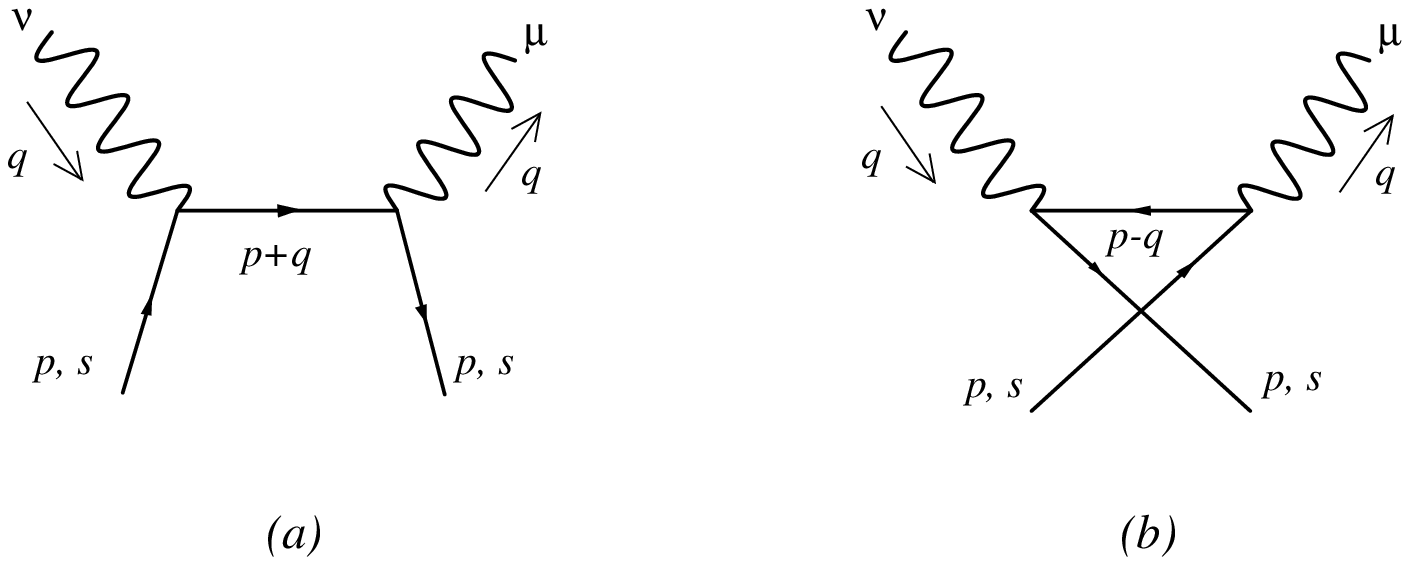}
\centerline{\tenrm FIGURE 7.}
{\tenrm \noindent The lowest order diagrams contributing to the quark matrix
element of the product of two electromagnetic currents.}
\endinsert
\eqn\melem{
\CM^{\mu\nu}=i\,\bar u(p,s)\,\gamma^\mu \,i{\slash p + \slash q \over (p+q)^2}
\,\gamma^\nu\, u(p,s) + i\,\bar u(p,s)\,\gamma^\nu \,i{\slash p - \slash q
\over (p-q)^2}
\,\gamma^\mu\, u(p,s).
}
Note that there is an overall factor of $i$ because we are computing $i$ times
the time ordered product in Eq.~\tmunu\ and there are no factors of $-ie$ at
the vertex. The crossed diagram (second term) can be obtained by the
replacement $\mu\leftrightarrow\nu$, $q\rightarrow-q$ from the direct diagram
(first term), so I will concentrate on simplifying the first term.
Expanding the denominator gives
\eqn\xiiithirteen{
(p+q)^2 =  2p\cdot q + q^2 = q^2\left(1 + {2p\cdot q\over q^2}\right) =
q^2\left(1-\omega\right),
}
since $p^2=0$ for an on-shell massless quark. The numerator can be simplified
using the $\gamma$ matrix identity
\eqn\gamiden{
\gamma^\mu\gamma^\alpha\gamma^\nu = g^{\mu\alpha} \gamma^\nu
+ g^{\nu\alpha} \gamma^{\mu} - g^{\mu\nu}\gamma^\alpha
+ i \epsilon^{\mu\nu\alpha\lambda}\gamma_\lambda\gamma_5,
}
\eqn\xiiifourteen{\eqalign{
\bar u(p,s)\gamma^\mu (\slash p + \slash q)
\gamma^\nu u(p,s)&=
\bar u(p,s)\Bigl[(p+q)^\mu\gamma^\nu +
(p+q)^\nu\gamma^\mu-g^{\mu\nu}\left(\slash p + \slash q\right) \cr
&\quad+ i
\epsilon^{\mu\nu\alpha\lambda}(p+q)_\alpha\gamma_\lambda\gamma_5\Bigr]
u(p,s).\cr
}}
For an on-shell massless quark,
\eqn\xiiififteen{
\slash p\ u(p,s)=0,\quad \bar u(p,s)\,\gamma_\lambda\,  u(p,s) = 2
p_\lambda,\quad \bar u(p,s)\,\gamma_\lambda \gamma_5\, u(p,s) = 2 h\,
p_\lambda,
}
where $h$ is the quark helicity. Thus the $\slash p$ and
$\epsilon^{\mu\nu\alpha\lambda} p_{\alpha}\gamma_\lambda\gamma_5$ terms both
give zero. (A note of caution: There have been many mistakes made by authors
trying to distinguish
$\bar u(p,s)\,\gamma_\lambda \gamma_5\, u(p,s)\, p_\sigma$ from
$\bar u(p,s)\,\gamma_\sigma \gamma_5\, u(p,s)\, p_\lambda$. They are equal
because of the equations of motion.)

Combining the various terms, and using
\eqn\xiiisixteen{
\left(1-\omega\right)^{-1} = \sum_{n=0}^{\infty} \omega^n,
}
gives
\eqn\firstterm{
\CM^{\mu\nu}_{\rm direct}=-{2\over q^2}\sum_{n=0}^\infty \omega^n
\left[ (p+q)^\mu p^\nu + (p+q)^\nu q^\mu - g^{\mu\nu}p \cdot q + i h
\,\epsilon^{\mu\nu\alpha\lambda}\,q_\alpha p_\lambda\right].
}
To complete the operator product expansion, we need the free quark matrix
element of the right hand side of the operator product.
The matrix element of the quark operators Eq.~\ov\ and \oa\
in a free quark state is
\eqn\ovmatrix{
\bra{p,s}\CO_V^{{\mu_1}\ldots {\mu_n}}\ket{p,s} = p^{\mu_1}\ldots p^{\mu_n},
}
\eqn\oamatrix{
\bra{p,s}\CO_A^{{\mu_1}\ldots {\mu_n}}\ket{p,s} = h\ p^{\mu_1}\ldots p^{\mu_n}.
}
The factors of $i$ and $2$ in Eqs.~\ov\ and \oa\ were chosen so
that no such factors appear in the matrix elements.

I will determine the coefficient functions for the spin
dependent term in the operator product expansion, and leave the
computation of the spin independent terms as a homework problem.
The spin dependent term on the left hand side of the operator product is
\eqn\spinterm{
\CM^{[\mu\nu]}=-{2\over q^2}\sum_{n=0}^\infty \omega^n\
  i h\, \epsilon^{\mu\nu\alpha\lambda}\,q_\alpha p_\lambda + \
(\mu\leftrightarrow\nu,\ q\rightarrow-q,\ \omega\rightarrow-\omega),
}
since $\omega$ is odd in $q$. The crossed diagram causes half the terms to
cancel, so that the matrix element is
\eqn\finalspinterm{\eqalign{
\CM^{[\mu\nu]}&=-{4\over q^2}\sum_{n=0,2,4}^\infty \omega^n
\ i h\, \epsilon^{\mu\nu\alpha\lambda}\, q_\alpha p_\lambda =
-{4\over q^2}\sum_{n=0,2,4}^\infty \left({-2 p\cdot q\over q^2}\right)^n
i h\, \epsilon^{\mu\nu\alpha\lambda}\,q_\alpha p_\lambda,\cr
&=\sum_{n=0,2,4}^\infty -{4\over q^2}\ {2^n q^{\mu_1}\ldots
q^{\mu_n}\over (-q^2)^n}\   i h\,
\epsilon^{\mu\nu\alpha\lambda}\,q_\alpha p_\lambda\ p_{\mu_1}\ldots
p_{\mu_{n}}, \cr
&=\sum_{n=0,2,4}^\infty 2\ {2^{n+1} q^{\mu_1}\ldots
q^{\mu_n}\over (-q^2)^{n+1}} \  i h\,
\epsilon^{\mu\nu \alpha\mu_{n+1}}\,q_\alpha\
p_{\mu_1}\ldots p_{\mu_{n+1}},\cr
}}
where the dummy index $\lambda$ has been replaced by $\mu_{n+1}$ in the
last line.
Replacing $n\rightarrow n+1$, and permuting the dummy indices $\mu_1 \dots
\mu_{n}$ gives
\eqn\mess{
\CM^{[\mu\nu]}=\sum_{n=1,3,5}^\infty 2\ {2^{n} q^{\mu_2}\ldots q^{\mu_n}\over
(-q^2)^{n}}\  i h\, \epsilon^{\mu\nu \alpha\mu_{1}}\,q_\alpha\ p_{\mu_1}\ldots
p_{\mu_{n}}.
}
The coefficient functions in the operator product depend only on $q$, and the
matrix elements depend only on $p$. We have separated the operator product into
pieces which depend only on $q$ and only on $p$. By comparing with
Eq.~\oamatrix, we can write Eq.~\mess\ as
\eqn\aaxialope{
\CM^{[\mu\nu]}=\sum_{n=1,3,5}^\infty 2\
{2^{n} q^{\mu_2}\ldots q^{\mu_n}\over (-q^2)^{n}}
\ i\epsilon^{\mu\nu \alpha\mu_{1}}\,q_\alpha \,\vev{O_A^{{\mu_1}\dots
{\mu_n}}},
}
so that
\eqn\axialope{
t^{[\mu\nu]}=\sum_{n=1,3,5}^\infty 2\ {2^{n} q^{\mu_2}
\ldots q^{\mu_n}\over (-q^2)^{n}}
\ i\epsilon^{\mu\nu \alpha\mu_{1}}\,q_\alpha \,O_A^{{\mu_1}\dots {\mu_n}}.
}
This is the operator product expansion for the spin dependent part of
$t_{\mu\nu}$.
We have omitted quark flavour types and quark charges in our analysis. The
correct answer for Eq.~\axialope\ has
$\CO_A$ replaced by $\sum \CQ_a^2 \CO_{A,a}$, where $\CQ_a$ is the
charge of flavour $a$ in units of $e$.  Odd spin axial vector operators
are charge conjugation even, and even spin axial vector operators are
charge conjugation odd. Since electroproduction is a charge conjugation
even process, only the odd spin axial vector operators appear in
Eq.~\axialope. I will leave the computation of the symmetric part of
$t_{\mu\nu}$ from Eq.~\firstterm\ as an exercise. The contribution can
be written in terms of only even spin vector operators. (Even spin vector
operators are charge conjugation even, and odd spin vector operators are
charge conjugation odd.)

The most general operator product expansion at twist two can be written in the
form
\eqn\ope{
\eqalign{
t_{\mu\nu} &= \sum^\infty_{n=2,4,\cdots}
\left(- g_{\mu\nu} +
{q_\mu q_\nu\over q^2} \right) {2^n q_{\mu_1}\ldots q_{\mu_n} \over \left( -
q^2\right)^n} \sum_{j} 2\ \cone\ O^{\mu_1\ldots \mu_n}_{j,V} \cr
&+\sum^\infty_{n=2,4,\ldots}
 \left( g_{\mu\mu_1} - {q_\mu
q_{\mu_1}\over q^2} \right) \left( g_{\nu\mu_2} - {q_\nu q_{\mu_2}\over
q^2}\right)
 {2^n q_{\mu_3} \ldots q_{\mu_n}\over \left( - q^2\right)^{n-1}}
\sum_{j} 2\ \ctwo \ O^{\mu_1\ldots \mu_n}_{j,V} \cr
&+ \sum^\infty_{n=1,3,\ldots}
 i\epsilon_{\mu\nu\lambda\mu_1}
q^\lambda\,  {2^n q_{\mu_2}\ldots q_{\mu_n}\over \left( - q^2\right)^n}
\sum_{j} 2\ \cthree\ O^{\mu_1\ldots \mu_n}_{j,A}\cr
&+\sum_{n=2,4,\cdots}^\infty  {2^n
q_{\mu_1}\ldots q_{\mu_n}\over \left(-q^2\right)^n}
\sum_{j}2\ \cfour \ O^{\mu\nu\mu_1\mu_2\mu_3\ldots\mu_n}_{j,\Delta}\
. \cr
}}
The sum in the various terms is over twist two quark and gluon operators. What
we have just shown is that $C_{j,n}^{(3)}$
at leading order is zero for the gluon operators, and is $\CQ_a^2$ for the
quark axial operator of flavour type $a$, as can be seen by comparing Eq.~\ope\
with Eq.~\axialope.
A similar calculation for the spin-independent pieces shows that at leading
order $C_n^{(1)}$ and $C_n^{(2)}$ are both equal to $\CQ_a^2$ for quark
operators of flavour type $a$, and all gluon operator coefficients are zero.
(The terms of type $C_n^{(4)}$ have only gluon contributions, and so vanish to
this order. They contribute to the structure function $\Delta(x)$ mentioned in
Sec.~12.)

To compute $T_{\mu\nu}$, we need the hadronic matrix element of the
operator product expansion, so we need to evaluate the hadronic matrix
elements of the operators Eqs.~\ov\ and \oa. The matrix elements of these
operators in the target are not known. We can parameterise the matrix elements
in terms of a known tensor structure times an unknown normalisation
coefficient. The matrix elements will be considered only in a spin-1/2 target.
For arbitrary spin targets, the tensor structure of the matrix elements is much
more complicated. The matrix elements of $\CO_V$ and $\CO_A$ can be written as
\eqn\zov{
\bra{p,s}\CO_V^{{\mu_1}\ldots {\mu_n}} \ket{p,s}=
V_n\ p^{\mu_1} \ldots p^{\mu_n},
}
\eqn\zoa{
\bra{p,s}\CO_A^{{\mu_1}\ldots {\mu_n}}\ket{p,s} =
A_n\ \CS\left[s^{\mu_1} \ldots p^{\mu_n}\right].
}
Using these matrix elements essentially undoes our free quark computation,
except for factors of $V_n$ and $A_n$, which were equal to one for the free
quark matrix elements. Once again, let me concentrate on the term antisymmetric
in $\mu\nu$. Substituting Eq.~\zoa\ into Eq.~\ope, one finds
\eqn\opesub{
T^{[\mu\nu]} = \sum_{n=1,3,5}^\infty 2\ C_n^{(3)}\ i
\epsilon^{\mu\nu\alpha{\mu_1}} q_\alpha\
{2^n q^{\mu_2}\ldots q^{\mu_n}\over (-q^2)^n}
A_n \ \CS\left[s^{\mu_1} \ldots p^{\mu_n}\right].
}
Let us write
\eqn\mixindex{
\CS\left[s^{\mu_1} \ldots p^{\mu_n}\right]
= s^{\mu_1} \ldots p^{\mu_n} + \CR^{{\mu_1} \ldots{\mu_n}},
}
where
\eqn\resttensor{\eqalign{
\CR^{{\mu_1} \ldots{\mu_n}} &= \CS\left[s^{\mu_1} \ldots p^{\mu_n}\right]-
s^{\mu_1} \ldots p^{\mu_n}\cr
&= {1\over n}\left[s^{\mu_1} p^{\mu_2} \ldots p^{\mu_n} +
p^{\mu_1} s^{\mu_2} \ldots p^{\mu_n} + \ldots
p^{\mu_1} p^{\mu_2} \ldots s^{\mu_n}\right]-s^{\mu_1} \ldots p^{\mu_n}\cr
&= -{n-1\over n} s^{\mu_1} p^{\mu_2} \ldots p^{\mu_n} +
{1\over n} p^{\mu_1} s^{\mu_2} \ldots p^{\mu_n} +\ldots+{1\over n}
p^{\mu_1} p^{\mu_2} \ldots s^{\mu_n}.\cr
}}
The tensor $\CR^{{\mu_1} \ldots{\mu_n}}$ has no completely symmetric part, and
so has spin $n-1$ rather than spin $n$.
Thus the contribution of $\CR$ to deep inelastic scattering is twist three,
rather than twist two, {\sl even though it came from the matrix element of a
twist two operator}. This is the Wilczek-Wandzura contribution to $g_2$, as
will be discussed in Sec.~10. Omitting the $\CR$ term in Eq.~\mixindex\ for the
moment, and substituting into Eq.~\opesub, we find
\eqn\texpand{\eqalign{
T^{[\mu\nu]} &= \sum_{n=1,3,5}^\infty  -{4\over q^2}\ C_n^{(3)} i
\epsilon^{\mu\nu\alpha{\mu_1}} q_\alpha s_{\mu_1}
 A_n \omega^{n-1},\cr
&= i \epsilon^{\mu\nu\alpha\lambda} q_\alpha s_\lambda\ {\tilde g_1\over p
\cdot q},\cr
}}
using the definition of $\tilde g_1$ from Eq.~\tstrfn. This gives us the power
series expansion
\eqn\goneexpand{\eqalign{
\tilde g_1 &= \sum_{n=1,3,5}^\infty-{4 p \cdot q\over q^2}\ C_n^{(3)}\, A_n
\,\omega^{n-1},\cr
\Rightarrow\quad\tilde g_1 &= \sum_{n=1,3,5}^\infty 2\,C_n^{(3)}\, A_n \omega^n
.\cr
}}
The operator product expansion has allowed us to compute $\tilde g_1$
as a power series in $\omega$ about $\omega=0$ in QCD. The $n^{\rm th}$
term in the power series is due to an operator of twist two and
spin $n$. The radius of convergence of the power series
is $\abs{\omega}=1$, since that is the location of the first singularity
in the  complex $\omega$ plane. This is precisely where the physical
region begins.

The structure functions can be computed near $\omega=0$, but not in the
physical region $1\le\abs{\omega}\le\infty$. To understand this better, it is
useful to use light-cone coordinates,
\eqn\lcone{
q^{\pm} = {1\over\sqrt{2}}\left(q^0\pm q^3\right).
}
Then the deep inelastic limit, $x$ fixed, $Q^2\rightarrow\infty$,
is the limit $q^+$ fixed, $q^-\rightarrow\infty$, because $Q^2=2 q^+ q^-$, and
\eqn\omegalc{
x={1\over\omega} = {q^+\over p^+}.
}
Thus deep inelastic scattering is not a short distance process; it probes the
structure of the hadron along the light-cone. However, if we look at the
unphysical region and let $\omega\rightarrow 0$, then $q^+\rightarrow \infty$
(and we already had $q^-\rightarrow\infty$). The expansion of the scattering
amplitude around $\omega=0$ is a short distance expansion, which is why one can
compute it in QCD.

We can relate $\tilde g_1$ in the unphysical region to its value in the
physical region using contour integration. The coefficient $2 C_n^{(3)} A_n$
can be extracted by the contour integral
\eqn\contint{
2 C_n^{(3)} A_n = {1\over 2\pi i}\oint_{\abs{\omega}=r<1}
\ \tilde g_1(\omega)\ {d\omega\over \omega^{n+1}},
}
over a circle in the complex $\omega$ plane around the origin, as shown in
\fig\contourfig{}. The contour of integration can then be deformed to the
contour $C$. This gives
\midinsert
\epsffile{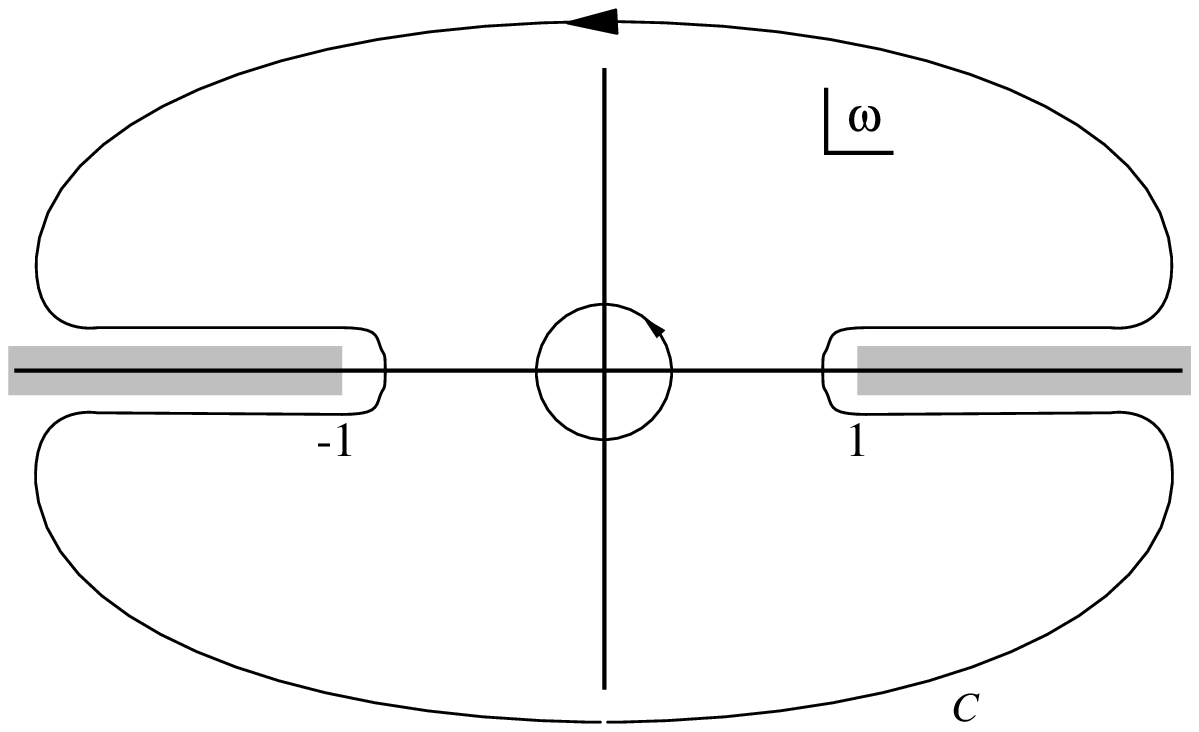}
\centerline{\tenrm FIGURE 8.}
\centerline{\tenrm \noindent The contours used to derive the moment sum rules.}
\endinsert
\eqn\continttwo{
2 C_n^{(3)} A_n
={1\over 2\pi i}\oint_{C}
\ \tilde g_1(\omega)\ {d\omega\over \omega^{n+1}}.
}
We now have to assume that the contour at infinity does not contribute to the
integral,
and can be neglected. In that case, only the discontinuities across the cuts
contribute to the integral over contour $C$. Using  the optical theorem,
Eq.~\optthm\
\eqn\xxyyfour{
\tilde g_1(\omega+i\epsilon)- \tilde g_1(\omega-i\epsilon) =
4\pi i\, g_1(\omega),
}
where $g_1(\omega)$ is the experimentally measurable structure function,
and
using Eq.~\evenoddrln
\eqn\wwwone{
g_1(-\omega)=-g_1(\omega),
}
we get
\eqn\goneint{
2 C_n^{(3)} A_n = 2\left[1 - (-1)^n\right]\int_1^\infty
g_1(\omega)\ {d\omega\over \omega^{n+1}},
}
where the $1$ term is from the right hand cut, and the $(-1)^n$ term is from
the left hand cut. Substituting $\omega=1/x$, and using the definition of a
moment, Eq.~\momdef,
gives
\eqn\gonesumrule{
2 M_n(g_1) = 2 \int_0^1 dx\ x^{n-1} g_1(x) = C_n^{(3)} A_n, \quad n\ {\rm odd}.
}
There is no sum rule for $n$ even, because Eq.~\goneint\ reduces to the
identity $0=0$.
The  sum rules for $F_1$ and $F_2$ are obtained in direct analogy to the
derivation of the $g_1$ sum rule. The result is that
\eqn\fsumrule{\eqalign{
2 M_n(F_1) &= C_n^{(1)} V_n, \quad n\ {\rm even},\cr
2 M_n\left({F_2\over 2x}\right) &= C_n^{(2)} V_n, \quad n\ {\rm even}.\cr
}}
The moment sum rules relate a quantity defined at high energy such as $F_1$ to
a low energy quantity, the zero momentum transfer matrix element of a local
operator.
One can derive sum rules in QCD only for the even moments of $F_1$, and
the odd moments of $g_1$. There are no sum rules for the other moments.
An attempt to
derive a sum rule for such a ``wrong moment'' leads to the
identity $0=0$, which is correct, but not very useful.
This is due to the symmetry properties of the structure functions under
charge conjugation (or crossing), Eq.~\tcrossinv. The Gottfried sum
rule discussed in Sec.~8 is an attempt to derive a sum rule for
the first moment of $F_1$, and is not true in QCD. It tries to relate a
charge conjugation even structure function $F_1$ to a charge conjugation
odd quantity, the number of $u$ and $d$ quark in the proton.

The only questionable assumption in the derivation of the sum rules is
that
the contour at infinity does not contribute to the integral. It should
be possible to tell experimentally whether this is true. If the contour at
infinity is important, then the sum rules will diverge near $x=0$, which can be
checked by measuring the small $x$ behaviour of the structure functions. The
sum rules become more convergent for the higher moments, so any problems with
convergence will only occur for the lowest few moments. There is no evidence
for any convergence problems for the sum rules in Eqs.~\gonesumrule\ and
\fsumrule.

\newsec{Applications of the Moment Sum Rules}

The previous section was a long and involved discussion of the derivation of
the moment sum rules from QCD. In this section, I would like to discuss some of
the consequences of the sum rules.
\bigskip
\leftline{{\it \underbar{$F_L$ in QCD}}}
\medskip
The sum rules for $F_1$ and $F_2$ imply that the moments of the longitudinal
structure function $F_L=F_2/2x - F_1$ are given by
\eqn\fflongsum{
2 M_n(F_L) = \left[ C_n^{(2)} - C_n^{(1)} \right] V_n, \quad n\ {\rm even}.
}
We have already seen that at lowest order, $C_n^{(1)}=C^{(2)}_n$. This implies
that $F_L=0$ at leading order in QCD. This is the Callan-Gross relation that we
obtained previously in Sec.~8 using the parton model. $F_L$ is more interesting
in QCD than in the parton model. At order $\alpha_s$,
$C_n^{(1)}\not= C^{(2)}_n$, so $F_L \not=0$, but is suppressed relative to
$F_1$ by $\alpha_s(Q)$. The same operator matrix elements $V_n$ occur in $F_L$
and $F_2$. The coefficients $C_n^{(1)}$ and $C^{(2)}_n$ can be computed in QCD
perturbation theory, so one can predict $F_L$ by using the measured value of
$F_1$ to determine $V_n$. This prediction agrees with experiment.
\bigskip
\leftline{{\it \underbar{The Ellis-Jaffe Sum Rule}}}
\medskip
Another interesting application is the sum rule for the first moment of $g_1$,
the Ellis-Jaffe sum rule,
\eqn\firstmom{
2 \int_0^1 g_1(x) dx = C_1^{(3)} A_1.
}
At one loop the coefficient is,
\eqn\xivone{
C_1^{(3)} = 1 - {\alpha_s(Q)\over\pi}.
}
The quantity $A_1$ is defined by the matrix element
\eqn\aonedef{
2 s^\mu A_1 = \bra{p,s} \left[{4\over 9} \bar u \gamma^\mu\gamma_5 u +{1\over
9} \bar d \gamma^\mu\gamma_5 d
+{1\over 9} \bar s \gamma^\mu\gamma_5 s\right]\ket{p,s}.
}
This is conventionally written as
\eqn\xivtwo{
A_1 = {4\over 9}\Delta u +{1\over 9}\Delta d
+{1\over 9}\Delta s,
}
where $\Delta q$ is defined by
\eqn\xivthree{
2 s^\mu \Delta q = \bra{p,s} \bar q \gamma^\mu\gamma_5 q
\ket{p,s}.
}
In terms of the parton distribution functions introduced in Sec.~8,
\eqn\xivfour{
\Delta q = \int_0^1 \left[q_+(x) - q_-(x) +\bar q_+(x) - \bar q_-(x)\right].
}
The notation is slightly confusing because $q_+(x) - q_-(x)$ is sometimes
referred to as $\Delta q(x)$.

The Ellis-Jaffe sum rule relates the first moment of the $g_1$ structure
function to the matrix elements of axial currents. This sum rule is special
because there is no twist two spin-one gauge invariant gluon operator, so the
right hand side of the sum rule only has the quark axial current operator, even
at higher order in QCD.
For $ep$ scattering, the linear combination $\Delta u -\Delta d$ in the proton
is equal to the axial decay constant $g_A$ in neutron beta decay by isospin
symmetry
\eqn\xivfive{
\bra{\rm proton} \bar u \gamma^\mu\gamma_5 u - \bar d \gamma^\mu\gamma_5 d
\ket{\rm proton} = \bra{\rm proton} \bar u \gamma^\mu\gamma_5 d \ket{\rm
neutron} = 2  g_A s^\mu.
}
The other non-singlet combination $\Delta u + \Delta d - 2 \Delta s$ is
determined by using $SU(3)$ symmetry and fitting to the $F$ and $D$
coefficients in hyperon semileptonic
decay,
\eqn\xivsix{
\Delta u + \Delta d - 2 \Delta s=3F-D,\quad g_A = F+D.
}
The Ellis-Jaffe sum rule can thus be written in the form
\eqn\ellis{
\int_0^1 g_1(x) dx = {C_1^{(3)}\over 18}\left[9F-D + 6 \Delta
s\right]=0.126\pm0.010\pm0.015,
}
where the experimentally measured value for the sum rule is determined by the
EMC
collaboration (Ref.~6).

The values of $F$ and $D$ can be obtained by fitting to the hyperon
semileptonic data. The standard results used in the literature are now out of
date, because the experimental results have changed significantly. I will use
the values
\eqn\fdvalues{
F = 0.47\pm0.04,\quad D=0.81\pm0.03,
}
from the paper by Jaffe and Manohar in Ref.~8. The EMC measurement, the
values of $F$ and $D$, and the Ellis-Jaffe sum rule
determine all three $\Delta q$'s,
\eqn\xiveight{
\Delta u = 0.74\pm0.10,\quad\Delta d = -0.54 \pm 0.10,\quad
\Delta s = -0.20 \pm 0.11,
}
and
\eqn\xivnine{
\Delta \Sigma \equiv \Delta u + \Delta d + \Delta s = 0.01 \pm 0.29.
}
Much of the recent interest in the spin structure of the proton is because of
the non-zero value for $\Delta s$, and the small value for $\Delta \Sigma$.

The structure function inequality Eq.~\fgineq\ implies
that $s(x) + \bar s(x) \ge \abs{\Delta s(x) + \Delta\bar s(x)}$. (This
can be proved by noting that the inequality holds for arbitrary values
of the electromagnetic charges of the quarks, and in particular holds
for $\CQ_u=\CQ_d=0$, $\CQ_s\not=0$.) The corresponding moment inequalities,
Eqs.~\foneineq\ and \frelgineq\ imply that the first moment
of $s(x) + \bar s(x)$, \ie\ the momentum fraction carried by $s$ quarks, bounds
all other moments of $s(x)+\bar s(x)$ and $\Delta s(x)+\bar \Delta s(x)$,
except the moment that appears in the Ellis-Jaffe sum rule.
The strange quark momentum fraction has been measured experimentally to be
0.026, so all other strange quark matrix elements should be small. The value of
$\Delta s$, Eq.~\xiveight, escapes this constraint.

The Bjorken sum rule is obtained by taking the first moment of the difference
$g_1^{ep}-g_1^{en}$ for proton and neutron targets. Using isospin invariance,
we see that
\eqn\bj{\eqalign{
\int_0^1 \left( g_1^{ep}(x) -g_1^{en}(x) \right)
dx &= {1\over 2}\, C_1^{(3)}\left[{4\over 9}\Delta u +{1\over 9}\Delta d
+{1\over 9}\Delta s\right]_{\rm proton -  neutron}\cr
&= {1\over 6}\, C_1^{(3)} \left(\Delta u - \Delta d\right)_{\rm proton}\cr
&= {1\over 6}\, C_1^{(3)} g_A.\cr
}}
Any flavour singlet contribution such as $\Delta s$ or gluons cancels
out in the Bjorken sum rule. This sum rule will  soon be tested
experimentally.

\vfill\break\eject
\bigskip
\leftline{\underbar{{\it The Wilczek-Wandzura contribution to $g_2$}}}
\medskip

Let us return to the tensor $\CR$ that we left out in the computation of $g_1$
in Sec.~9. Inserting $\CR$ into the formula for $T_{\mu\nu}$, Eq.~\opesub,
\eqn\gtwoexp{
T^{[\mu\nu]} = \sum_{n=1,3,5}^\infty -{(n-1)\over n}\left[s^{\mu_1} \,{2^n\,
 (p\cdot q)^{n-1}\over (-q^2)^n} - p^{\mu_1}\, {2^n\, (p\cdot q)^{n-2}\,
q\cdot s\over (-q^2)^n}\right] 2\, C_n^{(3)} A_n
\ i \epsilon^{\mu\nu\alpha{\mu_1}}q_\alpha
}
since each term with a $\mu_2 \ldots \mu_n$ index on $s$ contributes to the
second tensor structure. Comparing with the expansion Eq.~\tstrfn\
of
$T^{\mu\nu}$, we see that Eq.~\gtwoexp\ contributes to the structure function
$\tilde g_2$,
\eqn\gtwoform{\eqalign{
\tilde g_2(\omega) &= \sum_{n=1,3,5}^\infty 2 C_n^{(3)} A_n
\left[{1\over n} - 1\right]\omega^n\cr
&= -\tilde g_1(\omega) + \int_0^\omega \tilde g_1(\omega')\
{d\omega'\over\omega'}\cr
}}
using Eq.~\goneexpand\ for $\tilde g_1$. Taking the imaginary part of both
sides converts $\tilde g_{1,2}$ into the experimentally measurable structure
functions $g_{1,2}$. Since $\tilde g_1$
has an imaginary part only in the physical region, the integral over $\omega'$
can be restricted to the region $1\le\omega'\le\omega$. Changing the
integration variable to $x=1/\omega$ gives the Wilczek-Wandzura relation
\eqn\ww{
g_2(x) = -g_1(x) + \int_x^1 g_1(x')\ {dx'\over x'}.
}
This contribution to $g_2$ came from twist two operators,
but we have already seen that it is really a twist
three contribution because $\CR$ has spin $n-1$ instead of $n$.
There are also twist three operators in the operator product
expansion that contribute to $g_2$, so
the total contribution to $g_2$ can be written as
\eqn\xvione{
g_2(x) = \left[g_2(x)\right]_{\rm Wilczek-Wandzura}+
\left[g_2(x)\right]_{\rm twist\ three}.
}
{\sl Both terms are equally important, and there is no reason why the
Wilczek-Wandzura piece should be taken as a prediction for $g_2$.}

The Wilczek-Wandzura contribution arises because the twist two operator
Eq.~\oa\ is symmetric in all its indices. This weights the $p\cdot s$ term in
$T^{\mu\nu}$ by a factor of $(n-1)/n$ relative to the $q\cdot s$ term, where
$n$ is the spin of the operator. The standard decomposition of $T^{\mu\nu}$,
Eq.~\tstrfn\ weights the $p\cdot s$ and $q \cdot s$ terms by coefficients which
are independent of the power of $\omega$ in the structure function. It is this
mismatch which leads to twist two operators producing an effectively twist
three contribution to $g_2$.

\newsec{Anomalous Dimensions and Scaling Violation}

The $Q^2$ dependence of deep inelastic structure functions is an extremely
important subject, but I do not have the time to analyse it in any detail in
these lectures. I just want to give an outline of the QCD analysis. The
starting point is the moment sum rules for the structure functions,
\eqn\momone{
M_n(F(x)) \sim c_q\, \CO_q + c_g\, \CO_g,
}
in terms of quark and gluon operators, where $F$ stands for a generic structure
function. In deep inelastic scattering, the renormalisation scale $\mu$ is
chosen to be $Q$, to avoid large logarithms in the computation of the
coefficient functions. Thus Eq.~\momone\ is more properly written as
\eqn\momtwo{
M_n(F(x,Q^2)) \sim c_q(Q)\, \CO_q(Q) + c_g(Q)\, \CO_g(Q).
}
The $Q^2$ evolution of the structure functions is due to the $Q^2$ dependence
in the coefficients and operators in Eq.~\momtwo. The coefficient $c_q(Q^2) = 1
+ \CO(\alpha_s(Q))$ depends only weakly on $Q^2$.
The largest $Q^2$ dependence comes from operator renormalisation. The
renormalisation group equations for the operator can be written in the form
\eqn\xviione{
\mu{d\over d\mu}\ \CO^{\mu_1\ldots \mu_n} = -{g^2\over 16\pi^2}\,
\gamma_n\,
 \CO^{\mu_1\ldots \mu_n},
}
at one loop, where $\gamma_n$ is a calculable anomalous dimension.
Dividing by the coupling constant evolution equation
\eqn\xviitwo{
\mu{d\over d\mu}\ g = -b\, {g^3\over 16\pi^2},
}
and integrating, gives
\eqn\xviithree{
\CO(Q) = \left[{\alpha_s(Q)\over \alpha_s(Q_0)}\right]^{\gamma_n/2b} \CO(Q_0),
}
so we can relate the moment of the structure function at $Q_0$
to the moment at $Q$. This leads to a calculable $Q^2$
dependence of the structure functions. As $Q^2\rightarrow\infty$,
$\CO(Q)\rightarrow 0$, if $\gamma_n > 0$.

The actual situation is slightly more complicated. The
quark operator can be divided into singlet and non-singlet
pieces. The non-singlet pieces evolve as discussed above, but
the singlet pieces can mix with gluon operators. In the
singlet sector, one needs to solve a two
dimensional matrix renormalisation group equation. The computation is
similar
to the one sketched above, and the moment evolution equation is
a $2\times  2$ matrix equation in terms of the eigenvectors and eigenvalues of
the anomalous dimension matrix.

The $Q^2$ dependence of the Ellis-Jaffe sum rule is particularly simple.
There is no gluon operator that can contribute to the sum rule,
and thus there is no gluon operator which can mix with the singlet quark
currents. The non-singlet currents are conserved, and have no anomalous
dimension. The singlet current is not conserved because of the axial anomaly,
and has an anomalous dimension at two loops,
\eqn\xviifour{
\mu{d\over d\mu}\ j^\mu_{\rm singlet} = -N_f \left({g^2\over 4\pi^2}\right)^2,
}
where $N_f$ is the number of light flavours. Integrating the two loop equation
by dividing by Eq.~\xviitwo, gives
\eqn\xviifive{
j^\mu_{\rm singlet}(Q)
=
\exp\left[{N_f\over 8\pi}\left(\alpha(Q)-\alpha(Q_0)\right)\right]j^\mu_{\rm
singlet}(Q_0).
}
As $Q^2\rightarrow\infty$, $j^\mu(Q)$ approaches a finite value in contrast to
the case where the operator had a one loop anomalous dimension, Eq.~\xviithree.
The $Q^2$ variation due to the two loop anomalous dimension Eq.~\xviifive\
produces a very weak $Q^2$ dependence in $\Sigma$ in Eq.~\xivnine.

There are some constraints on the possible anomalous dimensions for the
structure functions in QCD. The unitarity inequality Eq.~\fgineq\ must
be true at all $Q^2$. This implies that the anomalous dimensions of the
axial vector operators must be larger than the vector operators, so that
for large $Q^2$, the $g_1$ structure functions falls off at least as
fast as $F_1$. Left and right handed quarks have the same QCD couplings.
This
implies that non-singlet axial and vector operators have the same
anomalous dimension, since they cannot mix with gluons. This is
not true for the singlet operators, because the Dirac trace in a
closed Fermi loop depends on whether or not there is an additional
$\gamma_5$ at the vertex. However, at one loop, it is still possible to show
that the $qq$ element of the singlet anomalous dimension matrix for vector and
axial operators is the same, since there are no closed Fermi loops in the
Feynman graph for the anomalous dimension.

\newsec{Outlook and Conclusions}
There were several new experiments on spin dependent deep inelastic
scattering discussed at this Winter Institute. The $g_1$
structure function of the proton will be measured more accurately, to
check the EMC measurement. The $g_1$ structure function of the
deuteron is also going to be measured. This, in combination with the
EMC measurement, will give the first measurement of $g_1$ for
the neutron, and should be able to test the Bjorken sum rule.
There are experiments at HERA to measure two new structure functions
which exist for spin one targets, $b_1(x)$ and $\Delta(x)$,
which I did not have time to discuss here. The new generation
of experiments are now sensitive enough to test the structure of the proton
that is not ``trivial,'' \ie\ effects from pair production such as the strange
quark distribution, Fermi statistics effects such as $\bar u \not=\bar d$, \etc

Let me briefly discuss the structure functions $b_1$ and $\Delta$ for
spin one targets. In terms of helicity amplitudes, the structure
function $b_1$ is proportion to
\eqn\bonefn{
b_1 \ \propto\ \left[  \CA(+,1;+1) + \CA(+,-1;+,-1) - 2 \CA(+,0;+0)\right],
}
and is the angular momentum two asymmetry in the cross-section for a
transverse photon to scatter off a polarised spin one target. The
structure function $b_1$ is a twist two structure function that can be
measured with a polarised target and an unpolarised beam. (All even
angular momentum structure functions can be measured this way; odd
angular momentum structure functions such as $g_1$ ($J=1$) need both a
polarised beam and polarised target.) The structure function
$\Delta(x)$ for a spin one target is proportional to
\eqn\deltarln{
\Delta\ \propto\ \CA(+,-1;-,+1),
}
and changes the helicity of the incident photon and target by two units.
It is also a twist two structure function, but gets no quark
contribution because the helicity of a quark cannot be changed by two
units. Thus the structure function is leading twist but order
$\alpha_s$, because it arises from gluon operators in the terms of type
$\cfour$ in Eq.~\ope. $\Delta(x)$ also vanishes for higher spin
targets that can be considered as a collection of non-interacting spin-1/2
nucleons, so
it should provide some interesting information on the structure of
higher spin targets.

I have tried to avoid any mention of models in this talk, except for
a brief discussion of the parton model in Sec.~9. Instead, I
concentrated on those aspects of deep inelastic scattering that
can be derived from QCD using only quantum field theory.
If a model calculation disagrees with experiment it is
the model which is wrong (or needs to be ``improved''); but
if a QCD field theory calculation disagrees with experiment, it is
the experimental result which is wrong, or it is evidence for a new fundamental
interaction.

Finally, I would like to thank Professor F.C.~Khanna for inviting me to the
Lake Louise Winter Institute. This work was supported in part by Department of
Energy grant
DOE-FG03-90ER40546, and by a Presidential Young Investigator award from the
National Science Foundation, PHY-8958081.
\vfill\break\eject
\newsec{References}

\noindent Here are a few references which might be useful for students
interested in learning more about the subject.

\frenchspacing
\parindent=1.0truecm
\bigskip
\item{1.} For an earlier review of polarised deep inelastic scattering:
\itemitem{}{V.W.~Hughes and J.~Kuti, {\it Ann. Rev. Nucl. Part. Sci.} {\bf 33}
(1983) 611.}
\bigskip
\item{2.} For a discussion of discrete symmetries such as $\CC$, $\CP$ and
$\CT$:
\itemitem{}{T.D.~Lee, {\it Particle Physics and Introduction to
Field Theory}, (Harwood Academic Publishers, Amsterdam, 1981).}
\bigskip
\item{3.} Some textbooks and lecture notes that discuss
the operator product expansion and deep inelastic scattering:
\itemitem{}{S.~Pokorski, {\it Gauge Field Theories}, (Cambridge
University Press, 1987)\semi
J.~Collins, {\it Renormalization}, (Cambridge University Press, 1984)\semi
Lectures by C. Callan and by D. Gross, in
{\it Methods in Field Theory}, ed. R.~Balian and J.~Zinn-Justin,
(North-Holland, Amsterdam, 1976).}
\bigskip
\item{4.} Some general references on QCD:
\itemitem{} R.D.~Field, {\it Applications of Perturbative QCD},
(Addison-Wesley, 1989)\semi
H.D.~Politzer, {\it Phys. Rep.} {\bf C14} (1974) 129.
\bigskip
\item{5.} The Bjorken and Ellis-Jaffe Sum Rules:
\itemitem{}{J.D.~Bjorken, \physrev{148}{1966}{1467}\semi
J.~Kuti and V.F.~Weisskopf, \physrev{D4}{1971}{3418}\semi
J.~Ellis and R.L.~Jaffe, \physrev{D9}{1974}{1444}.}
\bigskip

\item{6.} The EMC measurement of $g_1(x)$ for the proton:
\itemitem{}J.~Ashman, \etal, \pl{206}{1988}{364}\semi
J.~Ashman, \etal, \np{328}{1989}{1}.
\bigskip
\item{7.} Implications of the EMC results for the strangeness content of the
nucleon:
\itemitem{}S.~Brodsky, J.~Ellis, and M.~Karliner, \pl{206}{1988}{309}\semi
D.B.~Kaplan and A.V.~Manohar, \np{310}{1988}{507}.
\bigskip
\item{8.} The $g_1$ structure function, anomalies,  the gluon spin
and other controversial topics are discussed in:
\itemitem{}G.~Altarelli and G.G.~Ross, \pl{212}{1988}{39}\semi
R.D.~Carlitz, J.C. Collins, and A.H. Mueller, \pl{214}{1988}{229}\semi
R.L. Jaffe and A.V.~Manohar, \np{337}{1990}{509}\semi
A.V. Manohar, The $g_1$ Problem: Much Ado About Nothing, in
{\it Polarized Collider Workshop}, A.I.P. Conference
Proceedings No.~223 p. 90, ed. by J.~Collins, S.F.~Heppelman, and
R.W.~Robinett, (American Institute of Physics, New York 1991).
\bigskip
\item{9.} The computation of $F_L$ can be found in:
\itemitem{}A.~De R\'ujula, H.~Georgi, and H.D.~Politzer, \ap{103}{1977}{315}.
\bigskip
\item{10.} The coefficient functions and operator anomalous dimensions for spin
dependent scattering:
\itemitem{}M.A.~Ahmed and G.G.~Ross, \np{111}{1976}{441}\semi
J.~Kodaira, \np{165}{1979}{129}\semi
J.~Kodaira, S.~Matsuda, T.~Muta, T.~Uematsu, and K.~Sasaki, {\it Phys. Rev.}
{\bf D20} {(1979)} {627}\semi
J.~Kodaira, S.~Matsuda, K.~Sasaki, T.~Muta, and T.~Uematsu, {\it Nucl. Phys.}
{\bf B159} {(1979) }{99}.
\bigskip
\item{11.} Structure functions for higher spin targets, $b_1(x)$ and
$\Delta(x)$:
\itemitem{}P.~Hoodbhoy, R.L.~Jaffe, and A.V.~Manohar, \np{312}{1989}{571}\semi
R.L.~Jaffe and A.V.~Manohar, \np{321}{1989}{343}\semi
R.L.~Jaffe and A.V.~Manohar, \pl{223}{1989}{218}.
\bigskip
\item{12.} There has been a lot of recent work on $g_2$. See for example:
\itemitem{}E.V.~Shuryak and A.I.~Vainshtein, \np{201}{1982}{141}\semi
R.L.~Jaffe and X.~Ji, \physrev{D43}{1991}{724}.

\bye